\documentclass[draft,english,american,jgrga]{agujournal2019}
\usepackage[LGR,T1]{fontenc}
\usepackage[utf8]{inputenc}
\usepackage{array}
\usepackage{verbatim}
\usepackage{multirow}
\usepackage{tabularx}
\usepackage{graphicx}
\usepackage{rotfloat}

\makeatletter

\DeclareRobustCommand{\greektext}{%
  \fontencoding{LGR}\selectfont\def\encodingdefault{LGR}}
\DeclareRobustCommand{\textgreek}[1]{\leavevmode{\greektext #1}}

\usepackage{apacite}

\usepackage{txfonts}
\usepackage{graphicx}
\usepackage{aas-macros}
\usepackage{setspace}
\journalname{arXiv}

\makeatother

\usepackage{babel}
\begin{document}
\selectlanguage{english}%

\title{Modelling the long-term impacts of artificial warming on the Martian
water cycle and surface ice distribution.}

\authors{Ashwin S. Braude, \affil{1}\\
Edwin S. Kite, \affil{1,2}\\
Mark I. Richardson, \affil{3}\\
Alexandre Kling, \affil{1}\\
and Michael A. Mischna, \affil{4}}

\affiliation{1}{Astera Institute, Emeryville, CA, USA}

\affiliation{2}{Department of the Geophysical Sciences, University of Chicago, Chicago,
IL, USA}

\affiliation{3}{Aeolis Research, Chandler, AZ, USA.}

\affiliation{4}{Jet Propulsion Laboratory, California Institute of Technology, Pasadena,
CA, USA}

\begin{abstract}
Recent papers by Ansari et al. (2024, Science Advances 10, eadn4650)
and Richardson et al. (2025, arXiv eprint 2504.01455) have suggested
that global warming of the Martian surface (`terraforming') by 35
K to sustain local habitats above the melting point of water could
be achieved through the injection of engineered aerosols
into the Martian atmosphere. 
Using the MarsWRF 3D Global Climate Model, we investigate how artificial warming of Mars through engineered aerosol release would affect the planetary water cycle and the distribution of the major surface ice reservoirs. 
Within our model framework, every 20~K of global warming induces a tenfold increase in atmospheric water vapour content due to sublimation of H$_2$O ice from the North Polar Cap. This increases the potency of cloud radiative feedbacks which induces nighttime warming (\textasciitilde 5-10 K) at low latitudes,
but daytime cooling (up to 40 K) in the winter midlatitudes. Water is transferred from the edge of the North Polar Cap to the South Polar Cap and there is minor destabilisation of shallow northern midlatitude subsurface ice. As a result, seasonal sublimation of H$_2$O ice from the South Pole has an increased impact on the global water cycle. These
changes persist on Mars at least decades after loading
of the atmosphere with engineered aerosols ceases. Our model is limited by the gaps in our knowledge of present-day Martian weather and climate, and of the microphysics and radiative properties of candidate warming agents. Much more data is therefore needed before warming Mars could become feasible.
%Major uncertainties in our model results therefore
%motivate further research on the present-day Martian environment so
%that the consequences of large-scale alteration of the Martian climate
%system can be fully understood before it is implemented on a practical
%level.
\end{abstract}

\noindent \section{Introduction}

Although the Martian surface today is too cold and dry to sustain liquid water, overwhelming evidence shows that it was once warmer and wetter \cite<e.g.>{craddockhoward2002,hynek2010,kiteconway2024,mondro2025}, and may still be sufficiently warm in the subsurface for groundwater \cite{grimm2017}. Mars will naturally rewarm as the Sun brightens, but this will take billions of years and, if the water escapes into space, may result in a Mars that is even drier than it is today \cite{jakosky2024}. Previous studies \cite{mckay1991,zubrinmckay1997,debenedictis2025} have therefore assessed the feasibility of restoring a warm and wet Mars artificially using existing readily-available resources on the planet. Making Mars' environment more suitable for life faces many challenges including extreme cold, and a 6~mbar atmosphere that lacks oxygen and that fails to block damaging ultraviolet radiation. This is because much of Mars' ancient atmosphere was lost to space \cite{jakosky2018}, while known reserves of CO\textsubscript{2} ice on Mars could only double the present-day
atmospheric pressure \cite{jakoskyedwards2018,forget2024terraforming},
which is not sufficient to induce runaway warming \cite{Sagan1973,mckay1991}. 

Studies in the early 2000s \cite{gerstell2001,marinova2005,dicaire2013}
focussed on warming Mars by $>$30 K using a C$_3$F$_8$-dominated mix of
super-greenhouse gases, but this would require \emph{O}($10^{14}$) kg of fluorine - a scarce quantity in Mars soil \cite{forni2015,rampe2018role} - to produce. More recently, a range of alternative methods have been proposed to warm Mars, including through the solid-state greenhouse effect \cite{Wordsworth2019,wordsworthcockell2024} and with orbiting reflectors \cite{Handmer2024}. In this paper, we focus on the method proposed by \citeA{ansari2024} and \citeA{richardson2025} respectively (hereon referred to as ``A24'' and ``R25''), that involves releasing artificial `engineered' aerosols into the Martian atmosphere optimised to induce a powerful greenhouse effect by allowing sunlight to reach the surface, while redirecting upwelling thermal infrared radiation back to the surface. A24 and R25 suggest that these aerosols could warm Mars by >30 K in $\sim$2 Mars
years at current atmospheric pressure, with feedstock obtained either from the soil in the case of metal particles \cite{williams1979},
or from the electrolysis of Mars' CO\textsubscript{2} atmosphere \cite{hoffman2022moxie}.  This would be sufficient to sustain liquid water in the near-surface, which is one of several necessary prerequisites for microbes to thrive in Martian soil \cite{debenedictis2025}. 
%Practical implementation of the artificial warming of Mars may therefore not be beyond the reach of humanity in the near-future. On the flipside, this also increases the urgency of assessing and understanding the global impacts of artificial warming on the Mars climate system before it is carried out. 
On the other hand, it may not necessarily be desirable to warm Mars even if if it scientifically feasible to do so. Ethical arguments \cite<e.g.>{marshall1993,haqqmisra2012,schwartz2013,stoner2021,forget2024terraforming} against artificial warming include Mars' astrobiological and broader scientific value in its current virtually unmodified state. Creation of local habitats will therefore precede and take priority over global terraforming. Furthermore, secondary climate feedbacks on Mars could result in unforseen consequences on the Martian climate system that could counter any efforts to create habitable environments on Mars, or indeed even make Mars less habitable. This therefore motivates us to further study the effect of these climate feedbacks on an artificially warmed Mars, and to ensure that attempts to modify the Martian climate take into account the possibility of an `off-switch' \cite{kitewordsworth2025}, where
the planet can be restored to its pre-terraformed state if desired.

%The results of R25 indicated that the warming induced by these particles
%could be reversed by simply shutting off the release of these particles,
%at which point surface temperatures would return fully to their pre-terraformed
%state within just one or two Mars years. However, these studies neglected
%the effects of artificial aerosol release on changes to the Martian
%water cycle. 

Previous studies on this subject neglected a critical component of the Martian climate system which could potentially undermine efforts to increase its habitability: the response of the water cycle to artificial warming. Mars has $>$5 $\times$ $10^6$ km$^3$ water ice \cite{carrhead2015,jakoskyhallis2024},
stored in ice caps over both poles, and in shallow-buried midlatitude ice deposits \cite<reviewed in>{morgan2025} that may have formed during periods of high obliquity within the last million
years \cite{leightonmurray1966,Schorghoferforget2012,Aharonson2026}.  Mars' seasonal water cycle is presently dominated by the summer sublimation of the North Polar Cap. Each Mars year, $\sim$1~km$^3$ of H$_2$O sublimates from the North Polar ice Cap, which partly condenses out of the atmosphere as clouds, and is deposited seasonally onto the surface as frost that then resublimates later in the year (see \citeA{montmessin2017} for a review). By contrast, dust lag deposits and a permanent layer of CO\textsubscript{2} ice prevent water ice in the South Polar Cap from sublimating, which therefore only plays a minor role in the Martian seasonal 
water cycle. Simulations of recent Mars at
high obliquity \cite{haberle2003,levrard2004,mischna2003,forget2006,madeleine2009,madeleine2014,naarthesis}
show that warming the North Polar Cap would cause a) an increased concentration of water vapour in the atmosphere, and correspondingly more water-vapour greenhouse warming, b) more cloud formation, which can either heat or cool the surface \cite{madeleine2012,madeleine2014,kite2021}
and c) more seasonal surface frost \cite{bapst2015,lange2024}
at low latitudes, which would reflect more sunlight to space.
Changes to Mars'
water cycle through artificial warming could therefore warm or cool Mars
through complicated cloud radiative feedbacks \cite{haberle2013,madeleine2014,kite2021}. In addition,
simulations of the water cycle on early Mars show that warming can destabilise water ice reservoirs in the midlatitudes, which are valuable as a future source of propellant and drinking water, as well as being a scientific
record of climatic change in the last 0.1\%-1\% of Mars' history
\cite{bramson2021,becerra2021}. Studies of the early Martian climate suggest that a warmer Mars could result in an `icy highlands' scenario \cite{wordsworth2013,wordsworth2015}, where surface ice shifts to
high elevations. Here, the low atmospheric pressure would render the ice especially resistant to melting and sublimation, thereby making the return of the ice to its original location difficult, and hindering its use for human exploration.
Moreover, water ice is subject to negative feedbacks that make it resistant to warming. Firstly, ice can reflect more sunlight than bare ground, which helps to keep it cool and stabilise it against sublimation (``ice-albedo feedback''). Conversely, faster sublimation in a warmed climate would increase evaporitic cooling of warm ice, which can prevent surface melting \cite{Ingersoll1970,Schorghofer2020,khullerclow2024}.
Ice that has shifted in a warmed climate might therefore take decades or longer to revert to its original location if a decision was made to stop warming.
Understanding all of these effects on the water cycle induced by artificial warming is therefore of basic scientific interest.

Here, we use a 3-D Global Climate Model (GCM) to
assess the water cycle response to engineered Mars warming, including temperature feedbacks and possible ``tipping points'' in water ice distribution. In particular, we wish to address a) the effect of radiative cloud feedbacks on surface temperatures, b) changes in the distribution of surface and near-surface ice, and c) the reversibility of any of these changes on decade-long timescales. In section 2 we describe our
model and idealised warming agent. In section 3 we present our results on the effect of aerosol release on surface temperatures and atmospheric water concentration.
In section 4 we then present the results of our model when engineered
aerosol release is ceased, and how this would affect the planet's surface ice reservoirs.
Section 5 discusses data and model limitations that prevent the resolution of open questions about Mars warming,
and final conclusions are presented in Section 6.

\noindent \section{Model}

To model Mars' climate, we use the Mars version of the
Planet Weather and Research Forecasting (MarsWRF) 3-D GCM \cite{michalakes2004,Skamarock2005,richardson2007,toigo2012}. We run MarsWRF at 60 $\times$ 36 
spatial resolution (longitude $\times$ latitude), with 52 vertical layers parametrised in modified-sigma
coordinates from the surface to a top of atmosphere located at 
$\sim$120 km altitude to allow for an increase in the height of the Hadley Cell circulation induced by artificial warming. We do not model a thermosphere for the topmost layer of the atmosphere (spanning altitudes from 80 to 120 km) and instead assume the atmosphere to be well-mixed over all altitudes.
We use smoothed Martian topography (with a 5-point nearest neighbour
smoother at the resolution of the model) to minimise dynamical
instabilities in the model caused by sharp
topographic features such as Olympus Mons. We use a two-stream radiative transfer scheme with gas absorption modelled according to the correlated-k approximation \cite{mischna2012}. Outgoing longwave radiation is binned into seven individual spectral bands from 4.5 to 1000 microns, while incoming solar radiation is binned into seven spectral bands from 0.25 to 4.5 microns. Our model approximations result in net energy leakage into space of 2-3 W m\textsuperscript{-2} from the top of the atmosphere averaged over the Martian year. 
In these simulations, we
neglect both dust microphysics and dust-water ice coupling. Instead,
we prescribe seasonal changes in background dust optical depth according to the MY 36 non-Global Dust Storm scenario of \citeA{montabone2015,montabone2020}.
Water ice cloud particles are assumed to all be of equal radius, and form when the water vapour
content achieves 100\% saturation, with optical constants described
in \citeA{iwabuchiyang2011}. In our model, this size
is fixed at 5 \textgreek{μ}m effective radius, in line with observed values in the Aphelion Cloud Belt \cite{atwood2024}. Sublimation and deposition of surface water ice is modelled as described
in \citeA{lee2018}, with the albedo and emissivity of the surface only updated to the prescribed values for water ice if the amount
of water ice on the surface is greater than 5 g m\textsuperscript{-2}. 

We cap the surface temperature at the South Pole poleward of 85$^\circ$ S to be below the frost point of CO\textsubscript{2} in order to maintain a residual South Polar Cap for modern Mars conditions. We assume values of surface albedo and emissivity of 0.78 and 0.48 respectively for north polar CO\textsubscript{2} ice, and 0.43 and 0.747 respectively for south polar CO\textsubscript{2} ice, which were derived empirically through model tuning to atmospheric pressure variations observed by the Viking lander \cite{guo2009}. The albedo of south polar CO\textsubscript{2} ice is underestimated in our model relative to observations \cite<e.g.>{colaprete2005,schmidt2009,pommerol2011,garybicas2020}.  However, the behaviour of the residual South Polar Cap is also not fully understood and exhibits high interannual variability \cite{james2001,james2007}. As with most other models \cite<e.g.>{buhler2020}, a high CO\textsubscript{2} albedo in our model in line with observed values results in an overestimate of the amount of annual CO\textsubscript{2} ice deposition on and around the residual south polar cap. This, in turn, causes water ice to be too resistant to sublimation in summer, resulting in a high net annual accumulation of ice over the South Pole that is not in line with observations. The default properties of the South Polar CO\textsubscript{2} ice were therefore chosen to compromise between their observed values and how they affect the water and CO\textsubscript{2} cycle in our model. Nonetheless, in order to ascertain the sensitivity of the water cycle to the properties of the South Polar Cap under both control and warmed scenarios, we also ran an additional set of simulations in which temperatures at the South Pole were allowed to exceed the frost point of CO\textsubscript{2} and where emissivity of the South Polar Cap was decreased to 0.6. We ignore the effect of artificial heating not only on the change in albedo of the South Polar cap, but also on the destabilisation of buried CO\textsubscript{2} ice deposits there, which could potentially double Mars' atmospheric pressure \cite{buhler2020}. In practice, this release would be slowed by the latent heat of sublimation of CO\textsubscript{2} ice, with the full sublimation of the known CO\textsubscript{2} likely taking $>$100 years.

To initialise the model, we add a perennial North Polar ice cap down
to 80$^\circ$ N for latitudes between -180$^\circ$ and 115$^\circ$ E, and down to 85$^\circ$ N at
other longitudes (with an additional detached section of perennial
water ice between latitudes of longitudes of 120$^\circ$-180$^\circ$ E) to represent
the approximate size of the observed perennial cap. We parametrise
this cap as a quasi-infinite reservoir of water ice with an
assumed fixed albedo as a variable parameter - by default we use 0.4,
but we also perform model runs with 0.35 and 0.45 \cite{wilson2007,navarro2014}. We assume a
global mean surface pressure of 610 Pa and do not initialise a perennial
CO\textsubscript{2} or water ice cap in the South Pole. 

We initiate a total of nine `pre-runs' of the model until the global average water vapour column density
and surface ice distribution reaches annual steady state, which occurs
after $\sim$20 Mars years. Six of these pre-runs assume two
separate water radiative feedback scenarios and three North Polar Cap
albedo scenarios, These two water vapour feedback scenarios are a) where atmospheric water vapour is radiatively active but
water ice clouds are not; and b) where both are radiatively active. One pre-run is conducted with a North Polar Cap albedo of 0.4, where both water vapour and water ice are modelled as radiatively inactive. The final two pre-runs respectively correspond to radiatively active and inactive cloud scenarios when the South Polar CO\textsubscript{2} ice emissivity is set to 0.6. We
then start a new set of runs with the output from the completed pre-run, where
we inject engineered aerosol at a constant rate (in units of litres/s)
from a specific location on the surface of Mars for 20 Mars years. These runs are listed in Table \ref{table:listofruns}, and includes a 0 litre/s control where the pre-run continues for another 20 Mars years without any engineered aerosol injection.
We test the sensitivity of these runs to a) the location of the aerosol
release (from 0$^\circ$ N, 135$^\circ$ E by default, approximately corresponding to the coordinates of Gale Crater) and b) the dry depositional
velocity of particles in the lowest layer of the atmosphere (set
to 0.03 cm/s by default). Finally, we add two runs in which we inject aerosol at a constant rate for 5 Mars years, then cease aerosol release at the 5 Mars year mark (`shutoff') and allow the model to evolve for further 15 Mars years to study how quickly the planet reverts back to its pre-terraformed state.
%\begin{sidewaystable}
\begin{table}
\includegraphics[width=1\textwidth]{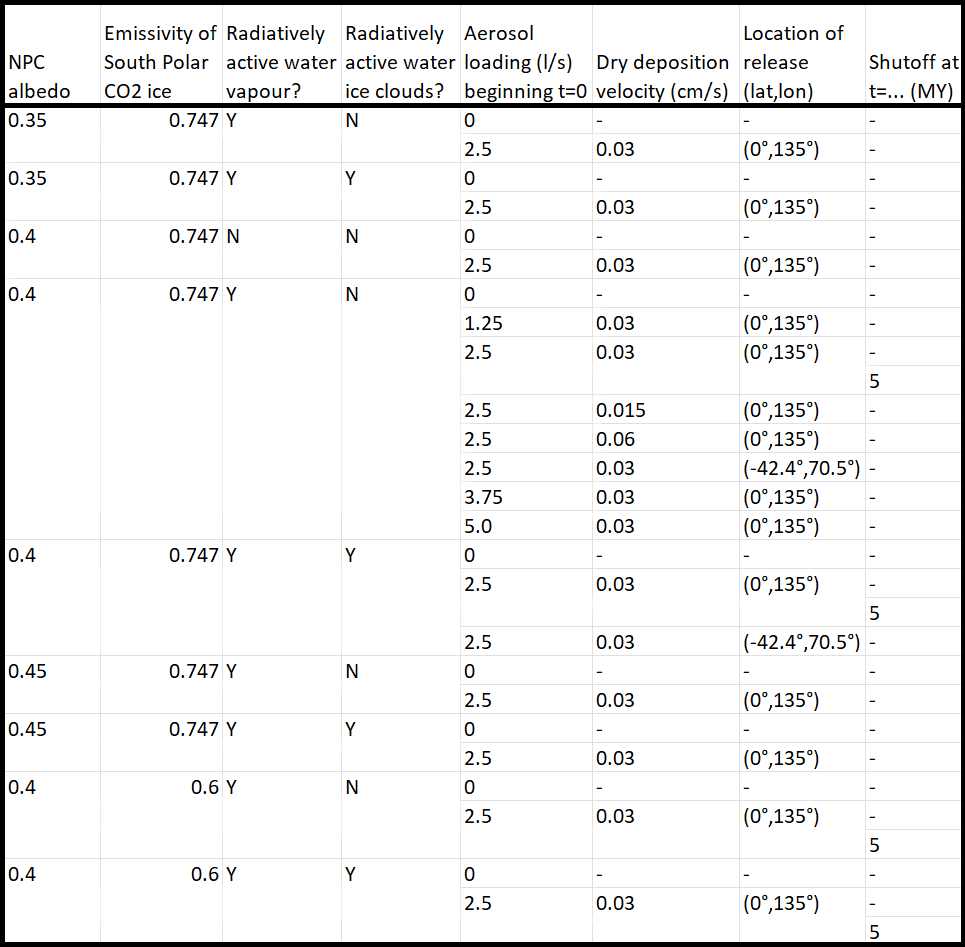}
\caption{List of model runs. Shared cells indicate
common model starting conditions. NPC = North Polar Cap. MY = Mars Years.}
%Default parameters (in bold) are
%assumed unless otherwise specified.
\label{table:listofruns}
%\end{sidewaystable}
\end{table}

A24 and R25 use finite-difference time-domain simulations to calculate the optical properties of a selection of candidate Mars-warming aerosols. However, none of these hypothetical particles have been shown to meet all the critera for deployment on Mars, including biocompatibility \cite{xuan2023}, manufacturability in a sustainable manner \cite{milliganelvis2019,kitewordsworth2025}, degradability in the Martian environment (R25), and benign interaction with spacecraft hardware \cite<e.g.>{farrell2004,anderson2009,barko2023}. We therefore use synthetic optical
constants for an idealised engineered aerosol as an input into our
model, as our goal is to model the effect of artificial warming on Mars' water cycle, not how the warming is produced.

\begin{figure}[h]
\includegraphics[width=1\textwidth]{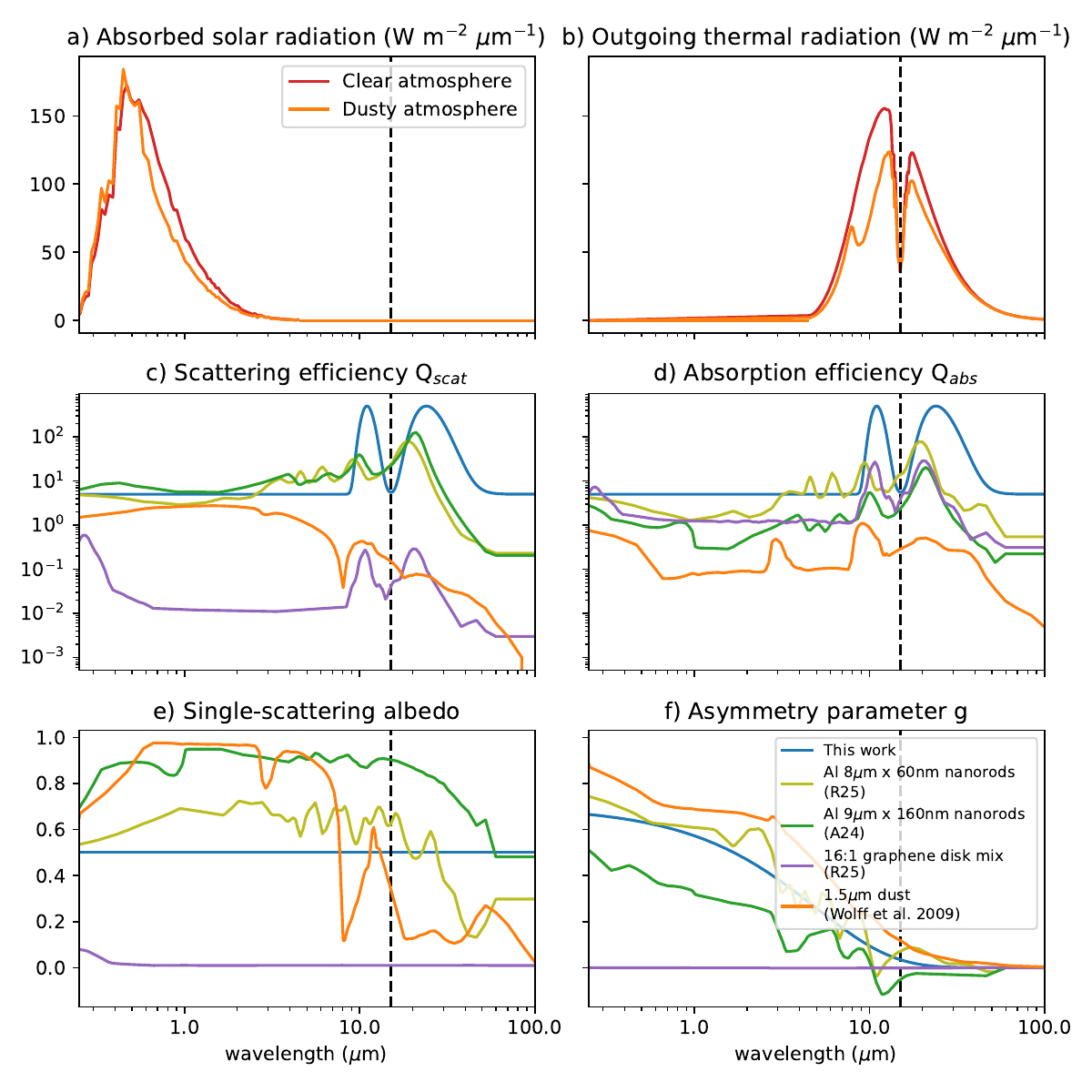}

\caption{Top row: a) simulated insolation reaching the surface and b) simulated emission at the top of the atmosphere (multiplied by a factor of 20) for both
clear and dusty conditions with optical depth \textgreek{τ} = 3, excluding
the effect of engineered aerosols.
  The vertical
black dashed line represents the 15 \textgreek{μ}m absorption band, where 
CO\protect\textsubscript{2} absorbs efficiently \cite{kutepov2025}. An engineered
aerosol absorbing at 15 \textgreek{μ}m will have relatively little additional warming effect.
c)-f): idealised Mars-warming aerosol optical properties, chosen to forward-scatter solar radiation to the surface while
blocking surface thermal emission. We also show the spectrum of Martian dust with an
effective radius of 1.5 \textgreek{μ}m \cite{wolff2009, haberle2019amesmodel} for comparison. Natural Mars dust lowers Mars dayside temperatures \cite{Streeter2020}, and is therefore not a suitable warming agent.}

\label{fig:greyaerosolspectrum}
\end{figure}

We parametrise the optical properties of this idealised aerosol using three variables as a function of wavelength from 0.2
to 100 \textgreek{μ}m: absorption efficiency $Q_{abs}$ (defined
as the absorption cross-section of the aerosol divided by its geometric
cross-section), scattering efficiency $Q_{scat}$ (defined
as the scattering cross-section of the aerosol divided by its geometric
cross-section), and an asymmetry parameter
$g=\{-1,1\}$ which is equal to the integral of the cosine of the scattering
phase function of the aerosol. Positive values of $g$
correspond to net forward-scattering, and an isotropic scatterer
has $g=0$. We set these parameters for
our idealised aerosol to share extinction trends with both the Al particle values tabulated in A24 and R25, which isotropically scatter thermal emission
from the Martian surface, and the graphene-disk values in R25, which correspond to near-zero scattering, high transparency to incoming solar radiation, and strong thermal absorption (Fig. \ref{fig:greyaerosolspectrum}).
We assumed a baseline extinction efficiency ($Q_{abs}$ + $Q_{scat}$)
equal to 10 across the wavelength range of analysis, and then added
two Gaussian absorption peaks with a peak value of $Q_{abs}$ + $Q_{scat}$ = 1000
centred around 11 \textgreek{μ}m and 24 \textgreek{μ}m respectively
and a variance of 2.5 \textgreek{μ}m (FWHM $\approx$ 3.7 \textgreek{μ}m)
to absorb the parts of the thermal emission spectrum not already absorbed
by atmospheric CO\textsubscript{2} (which has a strong absorption
peak at 15 \textgreek{μ}m). We assumed the particle was equal
parts absorbing and scattering with a single-scattering albedo $\left(\frac{Q_{scat}}{Q_{scat}+Q_{abs}}\right)$
equal to 0.5 at all wavelengths. Finally, we assumed a value of $g$
to follow exponential decay with wavelength approximately corresponding
to the trend seen in the asymmetry parameter of 8 $\times$ 0.06 $\times$ 0.06 \textgreek{μ}m
Al particles (R25). We assume each aerosol particle to have a mass
density of 2700 kg m\textsuperscript{-3} and have dimensions of
8 $\times$ 0.06 $\times$ 0.06 \textgreek{μ}m, similar to the Al particles of R25.

\noindent \section{Climate feedbacks on a warmed Mars}

\noindent \subsection{Global trends}

\begin{figure}[h]
\includegraphics[width=1\textwidth,totalheight=0.8\textheight,keepaspectratio,page=1]{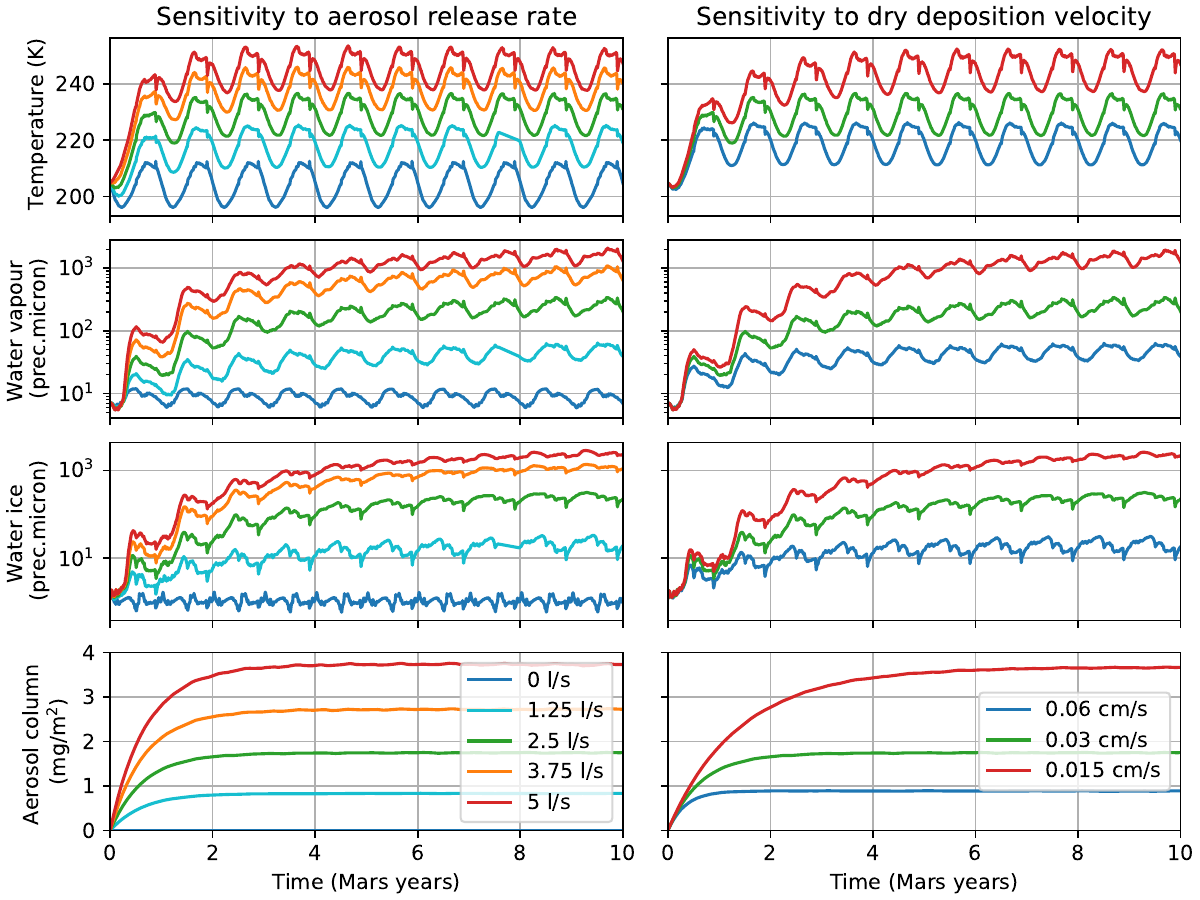}

\caption{Evolution over time of, from top to bottom: global average surface temperature,
atmospheric water vapour column depth, atmospheric water ice column
depth and aerosol column depth. Results are shown as a function of (\emph{left}) engineered
aerosol release rate for a fixed dry deposition velocity of 0.03 cm/s, and (\emph{right}) dry deposition velocity for a fixed aerosol release rate of 2.5 l/s. The model runs
shown in this figure all correspond to scenarios with radiatively active
water vapour but non-radiatively active clouds}
%, as the model becomes
%unstable for very high aerosol loadings when radiatively active clouds
%are turned on due to strong local warming feedbacks.}
\label{fig:timeseries}
\end{figure}

In this section we focus on global average trends excluding radiative cloud feedbacks, as aerosol loadings above 2.5 l/s with radiatively active clouds lead to instabilities in the model. Under this scenario, global average temperatures approach steady state within $<$3 Mars years of aerosol release (Fig.~\ref{fig:timeseries}): atmospheric engineered aerosol concentrations stabilise
as the rate of removal of aerosols from the atmosphere becomes equal
to the rate of release of aerosols. Steady release of 1.25~l/s,
2.5~l/s and 3.75~l/s of idealised aerosol, respectively,
results in the global average surface temperature of the planet stabilising
at 15~K, 25~K and 35~K above the control value of 205~K. The idealised aerosol is more potent than a realistic aerosol, being $\sim$12 times more powerful than that of Al particles in R25 which are close to theoretical optima for simple shapes; our values of surface temperature with the water cycle not taken into account 
correspond to the results of R25 with a factor-of-12 decrease in aerosol
release rates. Aerosol release causes rapid global warming proportional
to the release rate. However, the aerosol lifetime, and hence the release
rate that is required to maintain global heating of the surface,
is very sensitive to aerosol dry deposition velocities in the lowest layer of the atmosphere. On Earth, Brownian diffusion of particles from the atmospheric surface layer into the surface is the main surface sink of sub-$\mu$m particles \cite{seinfeldatmospheric1998,Farmer2021}.
However, there is no data for the dry deposition rate of sub-$\mu$m particles over desert surfaces under unstable conditions.
For Mars, the dry deposition rate of sub-$\mu$m particles depends on
unknowns including a) horizontal
winds and turbulence close to the surface and b) surface type and roughness, and is therefore difficult
to estimate \emph{a priori} (but see \citeNP{Li2025} for estimates). The global average aerosol loading that can be sustained in the atmosphere is inversely proportional to the dry deposition velocity of these aerosols.
%, and also decreases the amount
%of time required for the atmosphere to reach steady state since the
%single largest sink of aerosols due to dry deposition would be shortly
%upon release from the surface before they have time to diffuse far
%from the release site. 
A doubling of the dry deposition rate of aerosols
would halve the warming for a given aerosol flux, with warming more focussed
near the aerosol release point.

\begin{figure}[h]
\includegraphics[width=1\textwidth,totalheight=0.8\textheight,keepaspectratio,page=2]{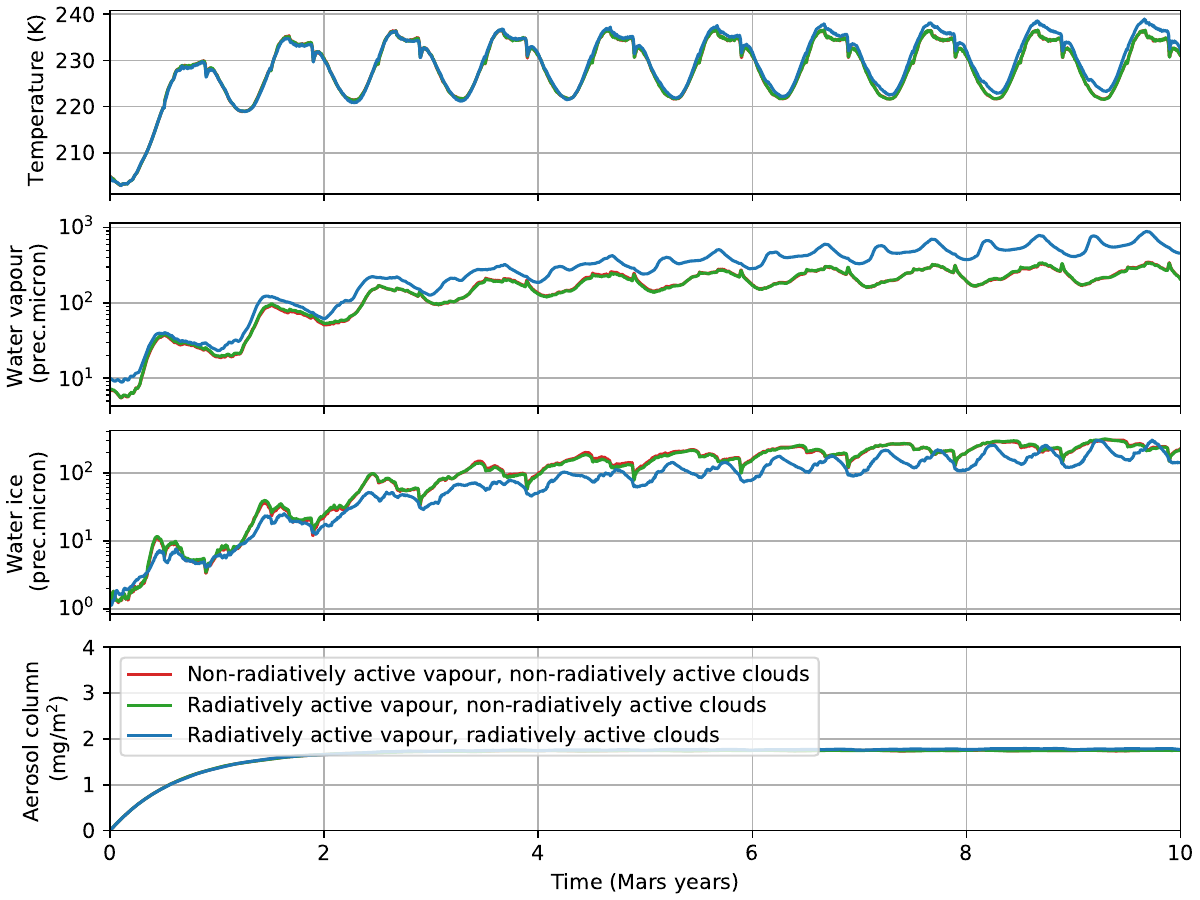}

\caption{Evolution of global average surface temperature, atmospheric water
vapour and ice column, and aerosol column depth for a 2.5 l/s aerosol release rate
scenario. Results are shown for three different radiatively active scenarios. The two scenarios with non-radiatively active clouds are mostly indistinguishable
from each other as the greenhouse effect from water vapour is weak.}

\label{fig:timeseriessens}
\end{figure}

Although the aerosol concentration reaches steady state
in $<$3 Mars years, water vapour continues to build up in the atmosphere for decades, which also results in greater cloud cover. In the first two Mars years of aerosol
release, this is due to increased summertime sublimation from the edge of the North Polar Cap, resulting in a tenfold increase in atmospheric water vapour for every 20 K of global average surface warming. Over time, this also increases winter deposition of frost in both the Northern and Southern midlatitudes, which then sublimates in summer. This acts to even out seasonal changes in atmospheric water vapour over the course of the year. As in \citeA{vos2025}, we find that the inclusion of a South Polar Cap which participates in the water cycle (which, in the case of \citeA{vos2025}, is due to the effects of high obliquity as opposed to global artificial warming) results in a shift in peak humidity towards perihelion, but where seasonal sublimation from the North Polar Cap still exceeds that of the South Polar Cap. Greenhouse warming from additional water vapour causes $<$0.1~K change to global average surface temperature. However, once the
global average amount of water ice cloud surpasses 100 precipitable
$\mu$m, warming due to radiative cloud feedbacks becomes significant (Fig. \ref{fig:timeseriessens}),
and global average surface temperatures increase long
after the concentration of engineered aerosols has stabilised.%
\begin{comment}
explain instability in model
\end{comment}
 We will elaborate on the climate impacts of this in the next subsection.

\begin{comment}
- buildup timescales of aerosol warming, compare warming wrt to aerosol
loading to other papers (unit conversion error in prec. $\mu$m plots?)

- sensitivity to dry deposition

- effect of water ice feedbacks on temperature buildup as a function
of aerosol loading (all in one plot?)
\end{comment}

\noindent \subsection{Effect of radiative cloud feedbacks on surface temperatures}

\begin{figure}[h]
\includegraphics[width=1\textwidth,totalheight=0.8\textheight,keepaspectratio,page=2]{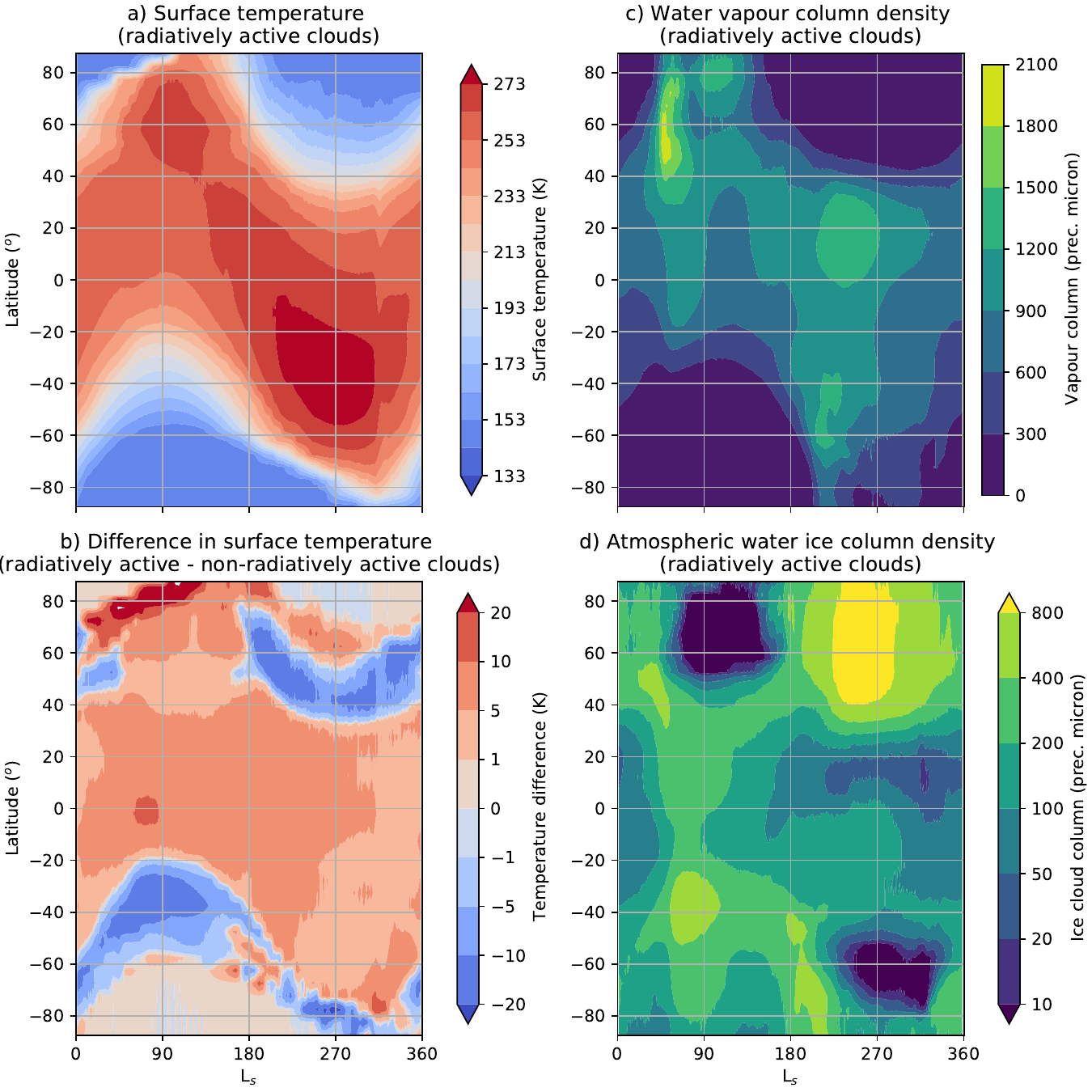}
\caption{\foreignlanguage{american}{Seasonal variations in (a) surface temperature and (b,d) the atmospheric water
cycle following 10 Mars years of aerosol release. (c) compares
 diurnally averaged temperature increase due to the direct versus indirect
effect of radiative cloud feedbacks in the model.}}

\label{fig:cloudfeedbackshovmoller}
\end{figure}

Adding radiative cloud feedbacks has major effects on the water
cycle (Fig.~\ref{fig:cloudfeedbackshovmoller}).
The moister atmosphere caused
by warming the North Polar Cap in turn results in strong cloud warming of the cap edge (which can affect the stability of the model if run over more than a decade), thereby further amplifying the seasonal sublimation of water ice from the North Polar Cap and increasing water ice deposition over the South Pole. This new South Polar
Cap participates in the water cycle analogously to the
North Polar Cap, resulting in a water cycle that is more hemispherically and seasonally
symmetric than today's cycle. This
includes a shift in Aphelion Cloud Belt timing \cite{clancy1996}
from a solar longitude (L\textsubscript{s}) =~90$^\circ$~-~140$^\circ$ to L\textsubscript{s}~=~60$^\circ$~-~100$^\circ$, as well as a new Perihelion
Cloud Belt over the equator at L\textsubscript{s}~=~220$^\circ$~-~270$^\circ$ caused by the movement of sublimated
water vapour away from the South Pole. The Perihelion Cloud Belt is at higher altitude than the Aphelion Cloud
Belt (\textasciitilde 40
km) because Mars is warmer at perihelion. %due to the warmer conditions present over the equator during perihelion.
%need numbers here? also maybe show the hovmollers for the non-ra case as well
In contrast to the Earth, where the greenhouse effect of water vapour is significant, the Martian atmosphere still remains dry compared to that of the Earth, and so the water vapour greenhouse effect remains minor. 
However, when radiatively active clouds are included in the
model, the clouds warm the atmosphere over the equator,
suppressing cloud formation at those altitudes and allowing water vapour to ascend up
into the thermosphere (>80 km altitude) during perihelion, as it does in simulations
of recent Mars at high obliquity \cite{gilli2025}. 
This likely accelerates atmospheric escape of water \cite{jakosky2018}, albeit
still negligibly slowly on human timescales and so is beyond the scope of this work.

%In our simulations, clouds form at very high altitudes over
%Northern Summer which restricts the movement of water vapour out of
%the atmosphere, although we caveat that neither supersaturation nor
%photochemistry were included in our model, and, in practice, a large
%proportion of the ascending water vapour molecules would likely photolyse
%at these altitudes before condensing to form clouds, which would also
%increase the rate of escape of water vapour out of the atmosphere,
%though the effect this has on the GEL of water in the Martian system
%would still be negligible on annual timescales \cite{jakosky2018}.
\begin{figure}[h]
\includegraphics[width=1\textwidth,totalheight=0.8\textheight,keepaspectratio,page=1]{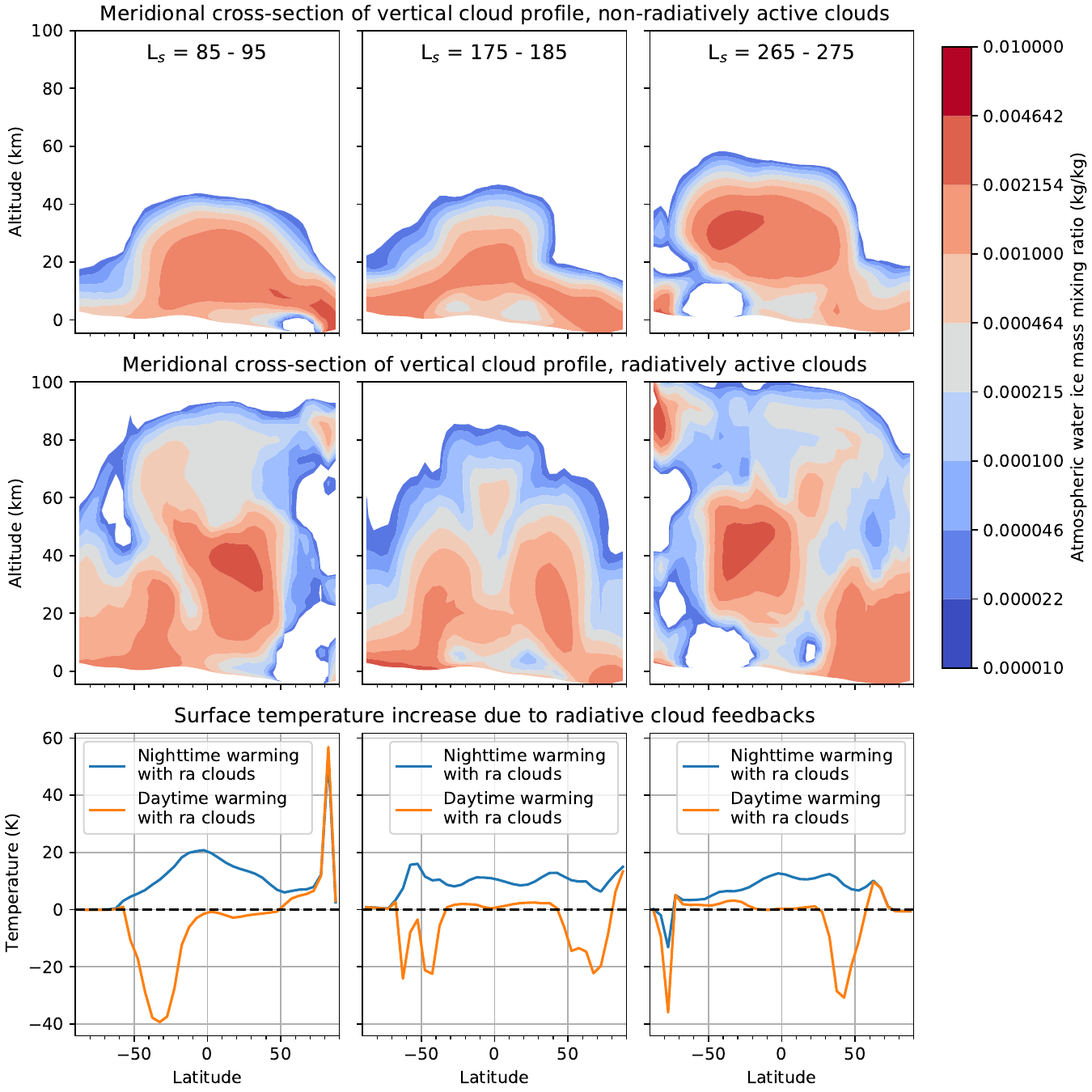}

\caption{Zonal average meridional cross-sections 10 Mars years after start of aerosol
release. Radiatively active ("ra") clouds warm the middle atmosphere,
allowing for water vapour to rise higher in the atmosphere before condensing. Cloud belts over
low- and midlatitudes therefore warm Mars during the night, while low-altitude clouds over the winter midlatitudes cause dramatic daytime
cooling. Recession of north polar CO\textsubscript{2} ice cap results in strong water ice sublimation in Northern summer, and hence strong warming due to radiative cloud feedbacks. }
\label{fig:diurnalcloudfeedback}
\end{figure}

\begin{figure}[h]
\includegraphics[width=1\textwidth,page=2]{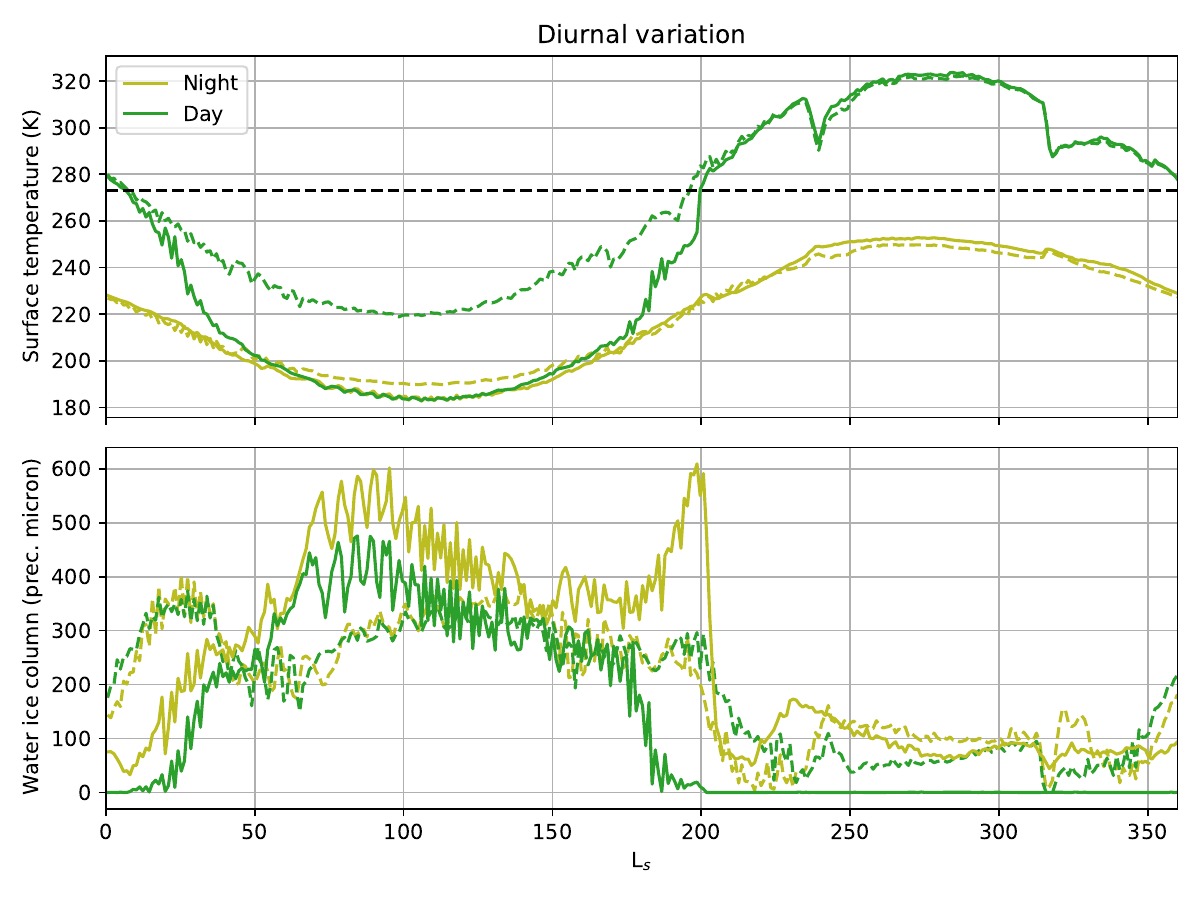}

\caption{\foreignlanguage{american}{Radiative cloud feedbacks (solid lines) over the Hellas Basin (42$^{\circ}$ S, 70$^{\circ}$ E) cause
dramatic daytime cooling in winter relative to when they are
excluded from the model (dashed lines). In summer, the Hellas
Basin warms by a few K during both day and night
due to radiative cloud feedbacks. The figure above corresponds to the climate 5 Mars
years after aerosol release starts.}}
\label{hellasplotsalbedo}
\end{figure}

\begin{figure}[h]
\includegraphics[width=1\textwidth,page=1]{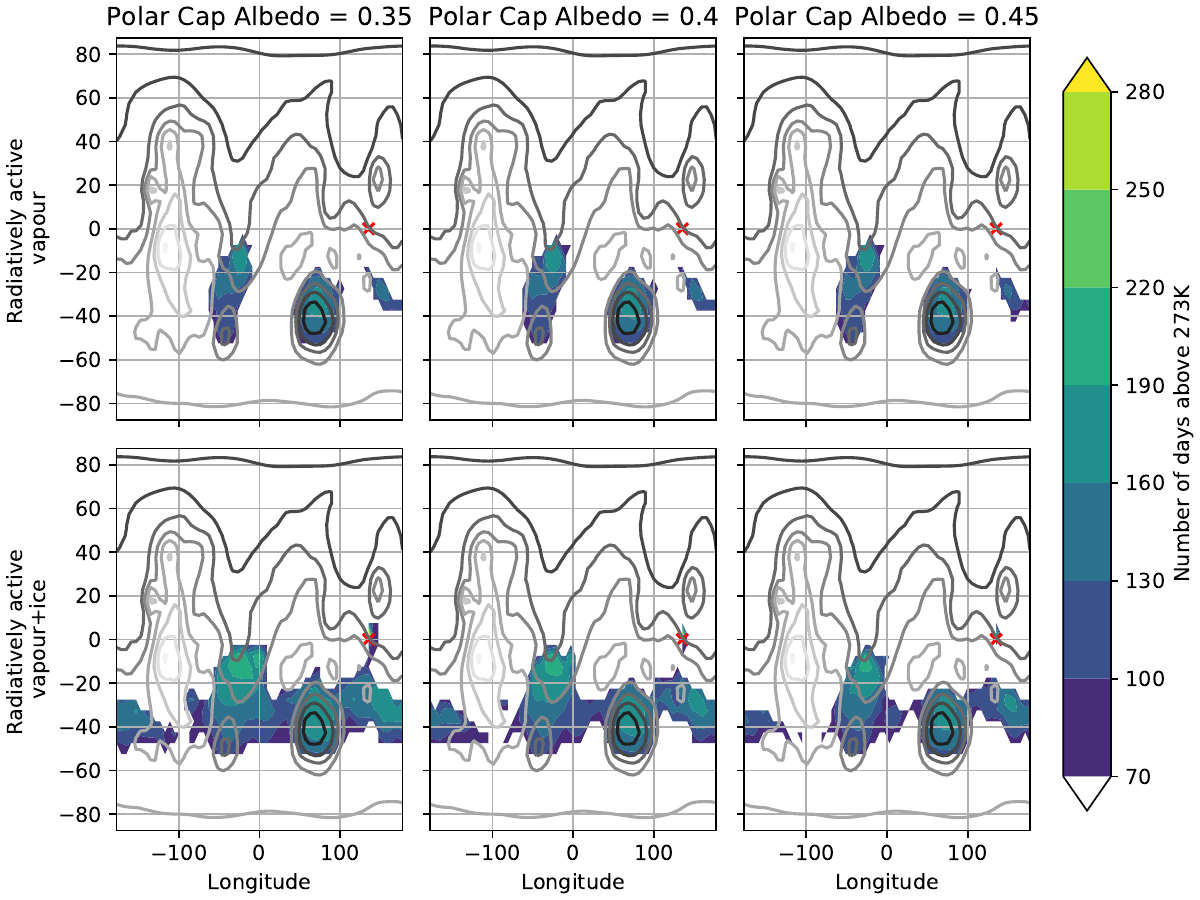}

\caption{\foreignlanguage{american}{As the North Polar Cap albedo decreases, more of the Southern midlatitudes experience
diurnal average surface temperatures above 273 K for more of the year,
thanks to the effect of enhanced radiative cloud feedbacks (bottom row). If the
effect of radiative cloud feedbacks are excluded (top row), warming sensitivity to albedo becomes negligible. The red cross
indicates the aerosol release site. Grey contour lines represent surface elevation in 2 km increments.}}
\label{fig:albedomap}
\end{figure}

Radiatively active clouds both warm and cool the surface depending on the latitude and time of day (Fig.~\ref{fig:diurnalcloudfeedback}). At night, clouds above 10 km altitude warm the low- and
midlatitudes by up to 20 K in most seasons. During the
day, thick clouds over the midlatitudes in the winter hemisphere 
cool the surface by up to 40 K. Frost forms at lower latitudes due to this cooling. In the Hellas
Basin (Fig. \ref{hellasplotsalbedo}), daytime cooling in winter can
reduce day-night cycles in temperature to $<$5~K
- remarkable given the low thermal inertia of the thin Martian
atmosphere. While clouds at these latitudes
cause colder winters, the \textasciitilde 5 K nighttime warming
in summer lengthens the amount of time with diurnally averaged temperatures above freezing
(Fig. \ref{fig:albedomap}). Hellas Basin warming is of particular interest in part because its relative high atmospheric pressure and shallow subsurface ice could increase
its potential as a microbial habitat, if warmed \cite{debenedictis2025}.

Comparison to 35$^\circ$ obliquity simulations in the literature shows similarities and differences. \citeA{madeleine2014} and \citeA{naarthesis} find \textasciitilde 20 K warming due to radiative cloud feedbacks
around the equator, with \citeA{madeleine2014} reporting globally averaged column densities of water vapour that are approximately equivalent to values we would expect for a 5 l/s aerosol loading scenario.
\citeA{naarthesis} finds even
greater (\textasciitilde 40-50 K) warming in the winter-hemisphere midlatitudes, where we observe
strong net cooling. \citeA{madeleine2014} find
this region to be a major zone of snowfall. 
This suggests model intercomparisons will be required to fully assess the effects of radiatively active clouds on Mars warming.

Changing the North Polar Cap albedo \cite{hale2005} causes only small changes in cloud feedback strength (Fig.
\ref{fig:albedomap}). A lower polar cap albedo causes greater water vapour sublimation in summer, and more winter cloud
condensation in the midlatitudes. The North Polar Cap might darken if sublimation of bright ice leads to accumulation of darker dust, if more dust falls on the cap due to a more vigorous dust cycle \cite{khuller2021},
or if engineered aerosol surface deposits are darker than the cap \cite{Sagan1973}. If polar darkening occurs, the atmosphere would moisten and might become cloudier.

\begin{comment}
- Diurnal effect of warming as function of latitude and time (two
figures: one Hovmoller, one meridional cross-sec)

- Sensitivity to albedo, number of days WITH DIURNAL AVERAGE above
273K (figure), include (sub)surface water ice map for Pioneer purposes.
focus on Hellas (figure). Latent heat due to subsurface ice not taken
into account. We only care about evaporitic cooling from subsurface
ice in warm season. Also want to check boiling during warm season
in Hellas, no geysers.

- How to include sensitivity wrt cloud particle size? Can increase
cloud particle size? Use 1.25 l instead of 2.5l
\end{comment}

\noindent \subsection{Regional warming of the Hellas Basin}

As liquid water could be most easily established in the Hellas Basin, it is natural to ask if engineered aerosol
warming can be confined there, as it could with orbiting reflectors, but not with artificial greenhouse gas emission. This would allow limited resources to be focussed
in areas where liquid water is most feasible while reducing the climate impacts elsewhere on Mars.

\begin{figure}[h]
\includegraphics[width=1\textwidth,totalheight=0.8\textheight,keepaspectratio,page=2]{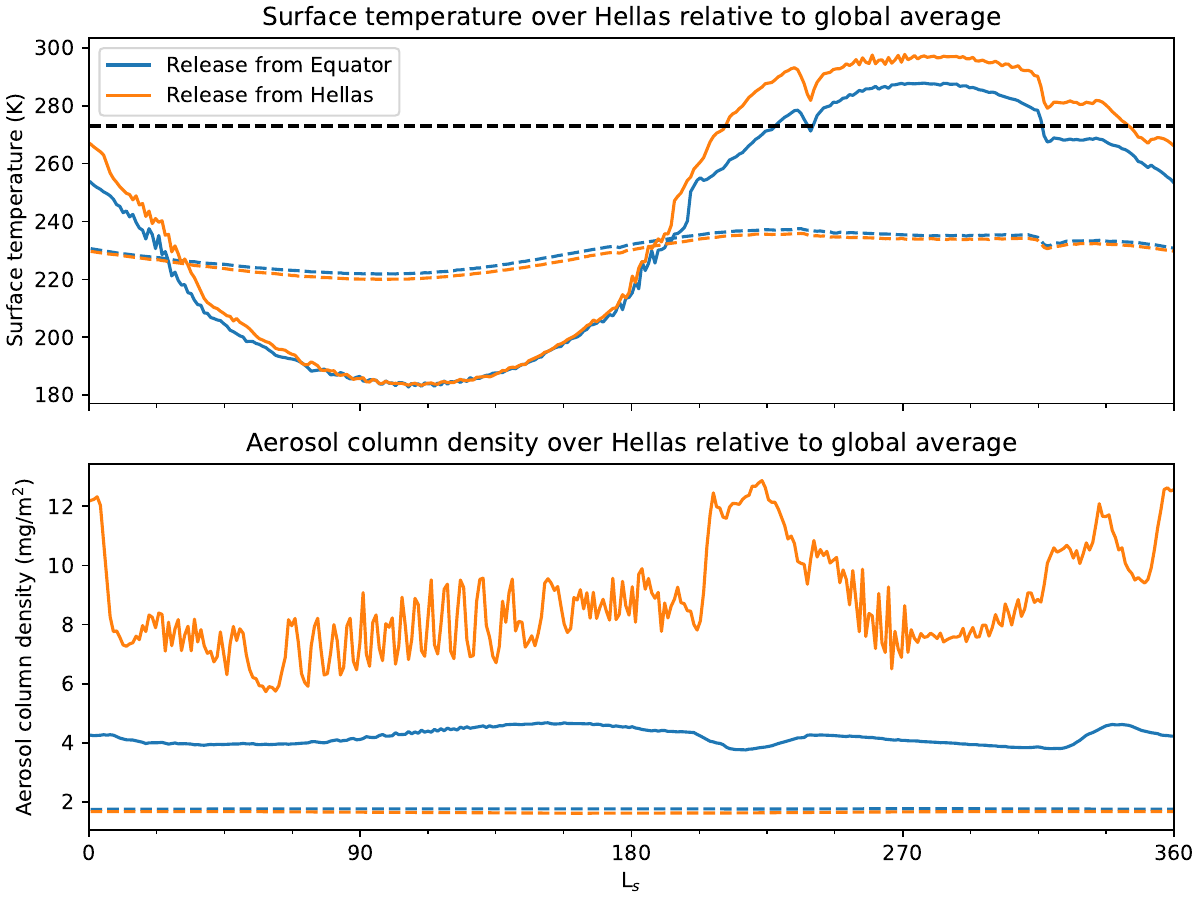}

\caption{\foreignlanguage{american}{Within the Hellas Basin (solid lines, 42$^\circ$ S, 70$^\circ$ E), warmer summers can occur with an aerosol release site within the Hellas Basin instead of a release site at the equator due to the increased localised concentration of aerosols (solid lines, bottom), even if the magnitude of the aerosol release is kept the same. This causes a 10\% reduction in global average aerosol column density
in the atmosphere (dashed lines, bottom) and a small decrease in global average temperatures (dashed lines,
top).}}
\label{fig:hellasconfineplots}
\end{figure}

\begin{figure}[h]
\includegraphics[width=1\textwidth,totalheight=0.8\textheight,keepaspectratio,page=1]{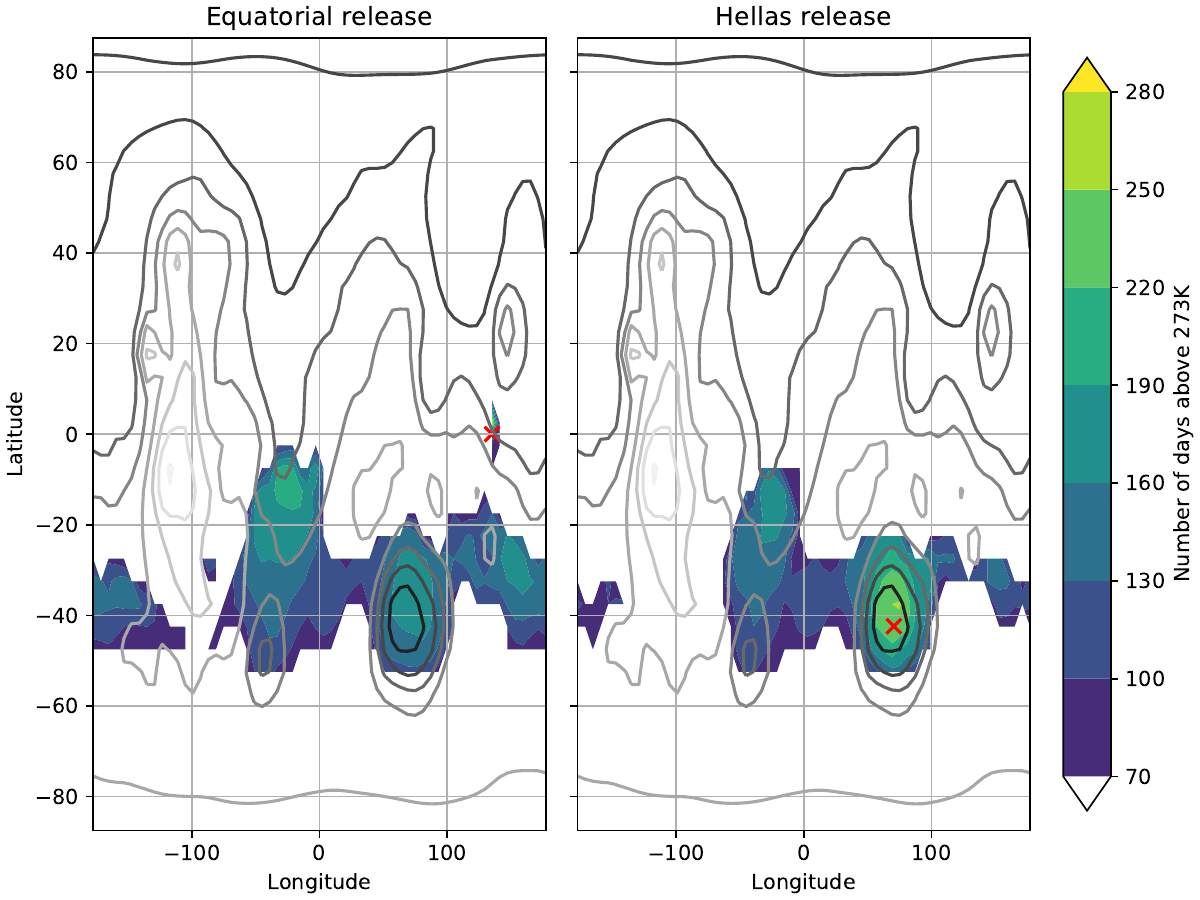}

\caption{\foreignlanguage{american}{Changing the aerosol release site (red crosses) from Gale
Crater (0$^\circ$ N, 135$^\circ$ E) to the centre of the Hellas Basin (42.4 $^\circ$ S, 70.5$^\circ$ E) lengthens the warm season with average temperatures above 273 K within the Hellas
Basin, while shortening the warm season elsewhere.} Grey contour lines represent surface elevation in 2 km increments.}
\label{hellasconfinemaps}
\end{figure}

We find that
releasing aerosols from the centre of the Hellas
Basin instead of from the equator raises average
summer temperatures in the Hellas Basin by 10~K (Fig. \ref{fig:hellasconfineplots}), despite a global reduction of aerosol density by 10\% due to the increased efficiency of
dry deposition in the Hellas Basin. Nonetheless, most of the aerosols disperse globally even if released from the bottom
of Hellas. Thus, while warming can be focussed in the region
of release, some places outside Hellas would
still have summer temperatures above 273~K if no other changes were made to the release
(Fig. \ref{hellasconfinemaps}), and global average temperatures would
only differ by 1-3 K. The idealised aerosols
studied here would therefore (for large release rates) have global effects regardless of where they were released. For real engineered aerosols, the lifetime of the aerosols in the atmosphere could be dramatically reduced by modifying particle size or density, which could confine warming.


\begin{comment}
- Compare heating in Hellas with nanoparticle loading when source of aerosol
release is changed (two figures?)
\end{comment}

\noindent \section{Reversibility and long-term impacts of artificially warming Mars}

\noindent \subsection{Effect of ceasing engineered aerosol release on the climate}

\begin{figure}[h]
\includegraphics[width=1\textwidth,page=1]{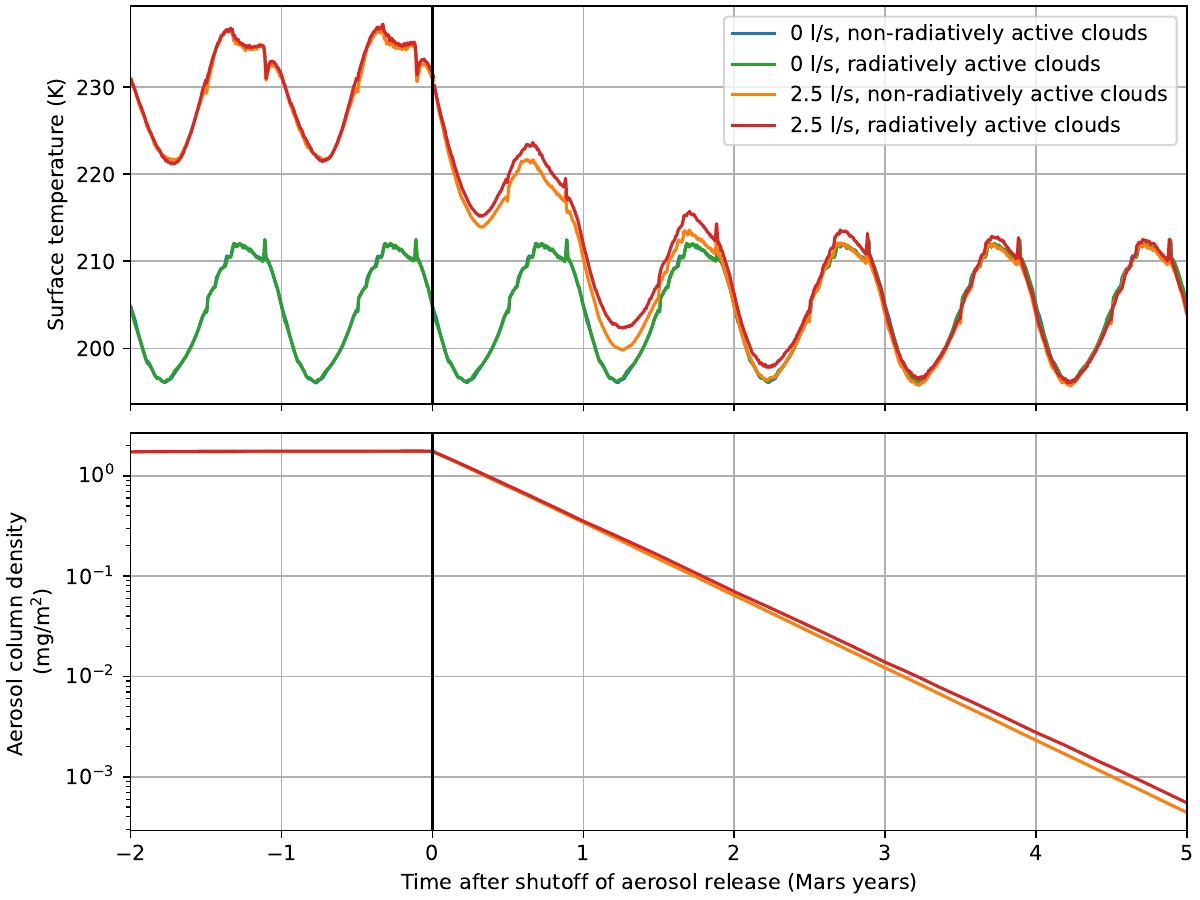}

\caption{\foreignlanguage{american}{Upon ceasing aerosol release 5 Mars years after start of 2.5 l/s aerosol
release, global average surface temperatures return to their pre-terraformed
values (green line) within just a few Mars years
as the aerosols are removed from the atmosphere on a 0.6 Mars year
e-folding timescale. Radiative cloud feedbacks have little effect
on aerosol removal timescales but slow down the reversal of aerosol warming.}}
\label{fig:warmingshutoff}
\end{figure}

\begin{figure}[h]
\includegraphics[width=1\textwidth]{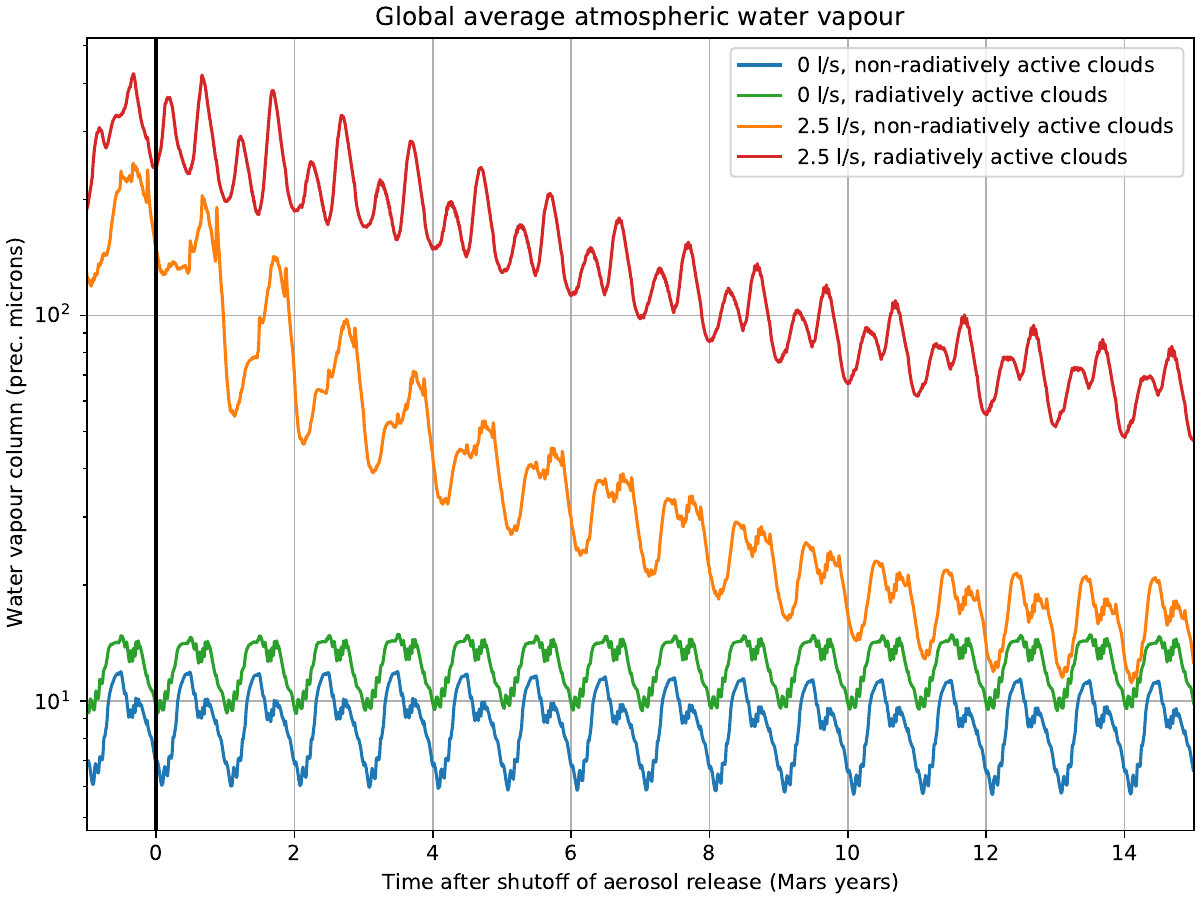}

\caption{\foreignlanguage{american}{Global average water vapour content in the atmosphere only gradually
declines following aerosol shutoff, eventually reaching a new equilibrium
level dictated by the formation of the new South Polar cap. The decline is even
slower if radiative cloud feedbacks are included as more water is stored seasonally in southern midlatitude frost.}}
\label{fig:vapourshuttofftimeseries}
\end{figure}

\begin{figure}[h]
\includegraphics[width=1\textwidth]{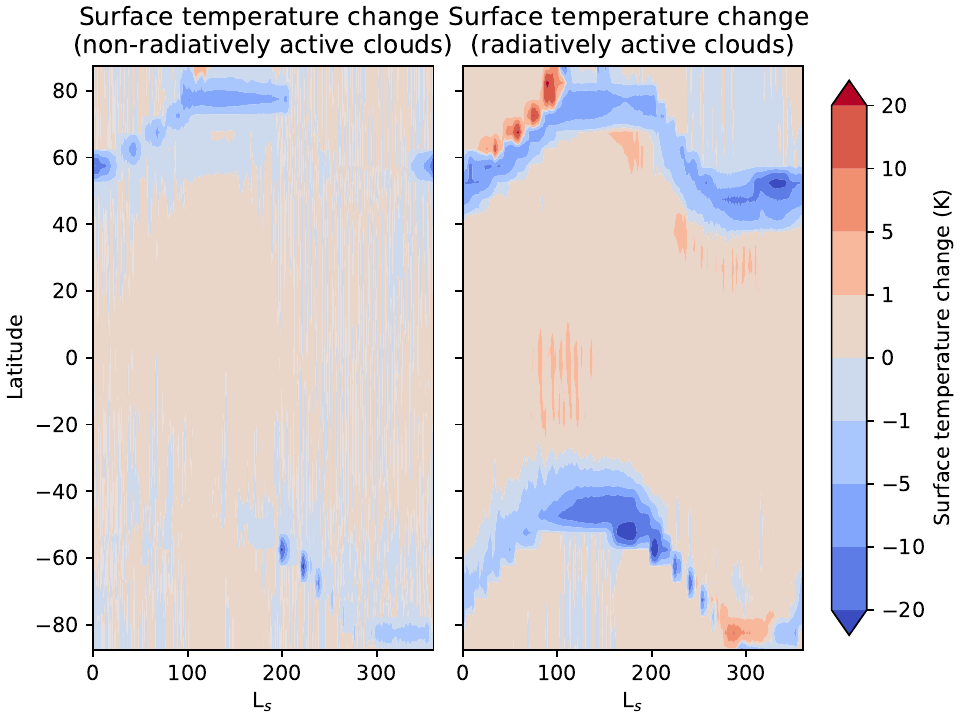}

\caption{\foreignlanguage{american}{Surface temperature increase 15 Mars years after aerosol shutoff relative
to the initial control case. Cloud-induced cooling in the winter midlatitudes persists for decades
after aerosol shutoff.}}
\label{fig:surftemphovmollershutoff}
\end{figure}

To test the reversibility of warming, we run two simulations (one with, and one
without, radiative cloud feedbacks). In both, we set 2.5 l/s release of idealised
aerosols for 5 Mars years, and then shut off the
release and let the model run for another 15 Mars years. The
idealised aerosols are removed by gravitational
settling and dry deposition with an e-folding timescale of $\sim$0.6 Mars years. As Mars' atmosphere remains thin, and its thermal
inertia low, the planet cools quickly,
and the warming directly induced by the aerosols
disappears after $<$4 Mars years (Fig. \ref{fig:warmingshutoff}).
However, the increased amount of ice cover means that the restoration of the Martian climate to pre-terraformed conditions is slow, taking at least decades, if not centuries (Fig.~\ref{fig:vapourshuttofftimeseries}). 
This is especially true when radiative cloud
feedbacks (which increase winter frost formation in the midlatitudes)
are included. While the warming
directly induced by the engineered aerosols disappears quickly, seasonal
temperature changes due to increased cloudiness will persist for much longer, with clouds causing $>$20 K cooling over the winter midlatitudes that persists 15 Mars years after aerosol release stops (Fig. \ref{fig:surftemphovmollershutoff}). %quantify

The size and duration of this change will depend on the duration of
the initial aerosol release: the earlier the aerosol
release ceases, the lower the bulk water vapour mass that would have
been sublimated away from the North Polar Cap, and hence the less permanent the
long-term impacts of the water cycle on seasonal temperature variation.

\subsection{Surface aerosol deposition}

\begin{figure}[h]
\includegraphics[width=1\textwidth]{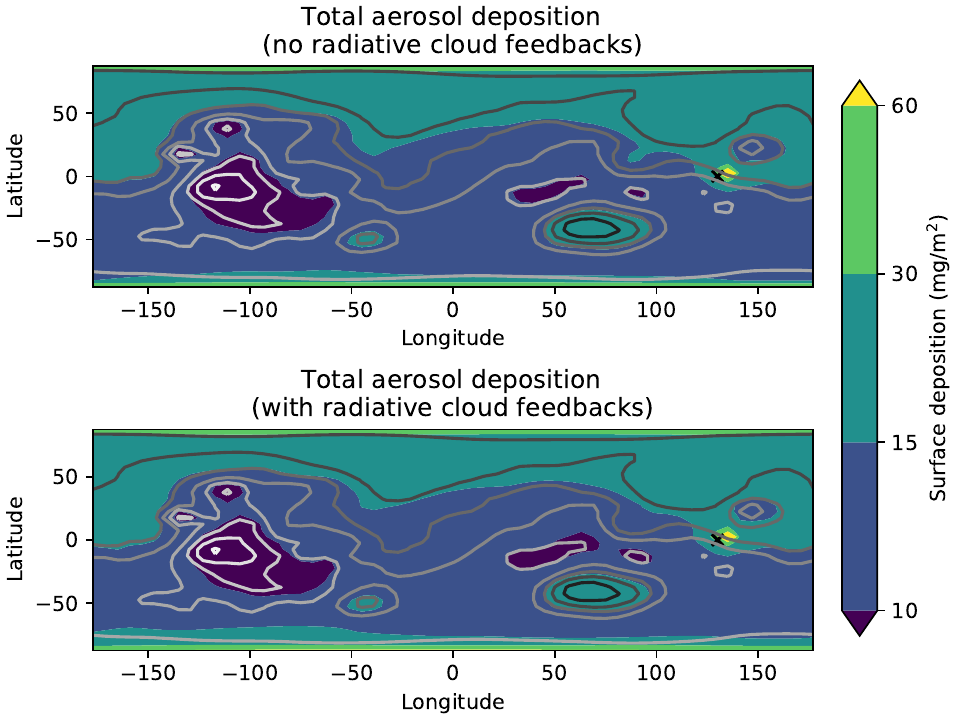}

\caption{\foreignlanguage{american}{Aerosol deposition on the surface following 5 Mars years of 2.5 l/s release
and 15 Mars years of shutoff. The distribution is comparatively uniform,
with more deposition at low elevations, high
latitudes and close to the release site (marked by the black cross). Including radiative cloud feedbacks makes little difference to the deposition.}}

\label{fig:nanoroddepmaps}
\end{figure}

Fig. \ref{fig:nanoroddepmaps} shows where engineered aerosols land on the surface. 
If we scale these results assuming our idealised aerosol has 12$\times$ greater warming per unit mass
than the Al particles in R25, then \textasciitilde 25 Mt of Al particles ( = \textasciitilde 170
mg m\textsuperscript{-2}) would be needed to produce the same amount
of warming over 5 Mars years as for the run shown in Fig. \ref{fig:nanoroddepmaps}%
\begin{comment}
check values
\end{comment}
, corresponding (for Al density of 2700 kg
m\textsuperscript{-3}) to a 
$\sim$13 nm global equivalent layer of particles deposited per year.
Aerosol deposition is near-uniform (whether or not radiative
cloud feedbacks are included), with a small increase in deposition at lower elevations. $\sim$2\% of the total aerosol
mass is deposited $<$500 km from the release site. Deposition is enhanced at the summer pole
 by the Hadley cell circulation acting to concentrate particles there. Over time, this could change polar cap albedo, and thus  the sublimation rate of water vapour into the atmosphere. %
\begin{comment}
compare with cospar planetary protection values. Is it actually enough
to induce any perceptible albedo change?
\end{comment}
In reality, some particles
would be re-lofted back into the atmosphere rather than staying on the surface,
and this could extend the lifetime of aerosol warming.
Alternatively, the particles may be designed to degrade into materials that are abundant on Mars. These processes would have to be quantified for real candidate warming agents using experiments in the laboratory.

\begin{comment}
- Timescales of warming ceasing (figure), particle deposition map (figure),
mention lofting
\end{comment}

\noindent \subsection{Effect of Mars warming on surface water ice}

\begin{figure}[h]
\includegraphics[width=1\textwidth]{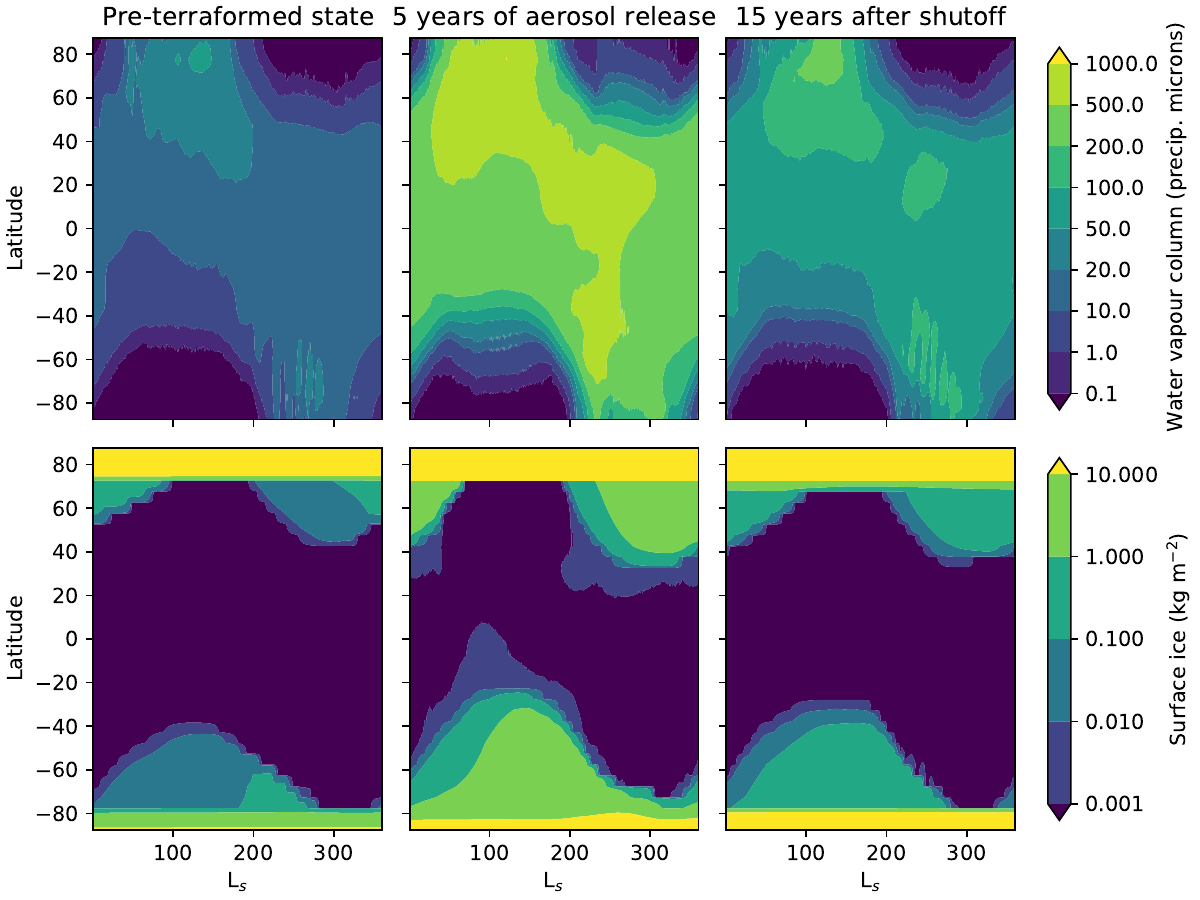}

\caption{\foreignlanguage{american}{Warming of the Martian atmosphere (centre) results in an
increase in atmospheric water vapour (top row) and thicker and more latitudinally-extensive frost in the winter midlatitudes (bottom row). Even 15 Mars years after
aerosol shutoff (right), the atmosphere remains moist relative to its pre-terraformed state, winter frost remains thicker, and a perennial South Polar water ice cap remains.}}
\label{fig:surfaceicehovmoller}
\end{figure}

The primary net source of water vapour in a warm Mars scenario is the region of perennial ice that, at the model resolution, spans latitudes between 75 and 80$^\circ$ N. This corresponds to a detached surface ice deposit spanning an area of approximately 100,000 sq km, separated from the main North Polar Layered Deposit by the Olympia Undae dunes region. We predict a net sublimation of 15-20 Gt/yr of water ice from this region, corresponding to \textasciitilde20 cm of the topmost layer of ice per Mars year. For comparison, if we assume a net yearly accumulation of water ice on the North Polar Cap within the 0.5-1 mm/yr range predicted by models \cite{landis2016,levrard2007,hvidberg2012}, this would correspond to the permanent loss of several centuries of the climate record stored in the ice deposits at 75$^{\circ}$ - 80$^{\circ}$ N every year.
Water vapour sublimating from the edge of the North Polar Cap migrates towards two (and potentially three) major sinks. Firstly, winter
frost thickens and spreads to lower latitudes as the atmosphere becomes more moist. This is aided by winter daytime cooling by clouds. Diurnally persistent frost in the
Southern hemisphere extends up to 20$^\circ$S during
aphelion season, while diurnally persistent frost in the Northern
hemisphere extends down to 30$^\circ$N. This compares to a maximum extent down
to 40$^\circ$ either side of the equator in the no-warming case
(Fig. \ref{fig:surfaceicehovmoller}). If we neglect the radiative effects of clouds, we also find that within approximately 7 Mars years of aerosol release, most of the excess water vapour added to the Martian system by the summer sublimation of the edge of the North Polar Cap is re-deposited over the Planum Boreum (>80$^\circ$ N), where temperatures remain cold for a larger proportion of the year and where a moister atmosphere favours net deposition of ice. This stabilises the global-average water vapour content of the atmosphere over the year. The inclusion of radiative cloud feedbacks, however, has the opposite effect, increasing summer temperatures over Planum Boreum and causing water vapour to sublimate from the region into to the Martian atmosphere. The contribution of the Planum Boreum to the climate system in a warm Mars scenario is therefore uncertain and highly dependent on model parametrisations of cloud microphysics (see Section 5). By contrast, the trace
amounts (<1 mm equivalent depth) of nighttime frost that forms atop low-latitude mountains - particularly Tharsis,
Syrtis Planum and Elysium Mons - sublimates at sunrise. There is thus no 
migration of water towards high-altitude cold traps as proposed for
Early Mars \cite{wordsworth2013,wordsworth2015}. This is because engineered aerosol warming is more effective at low latitudes than at the poles (R25). 

\begin{figure}
\includegraphics[width=0.45\textwidth]{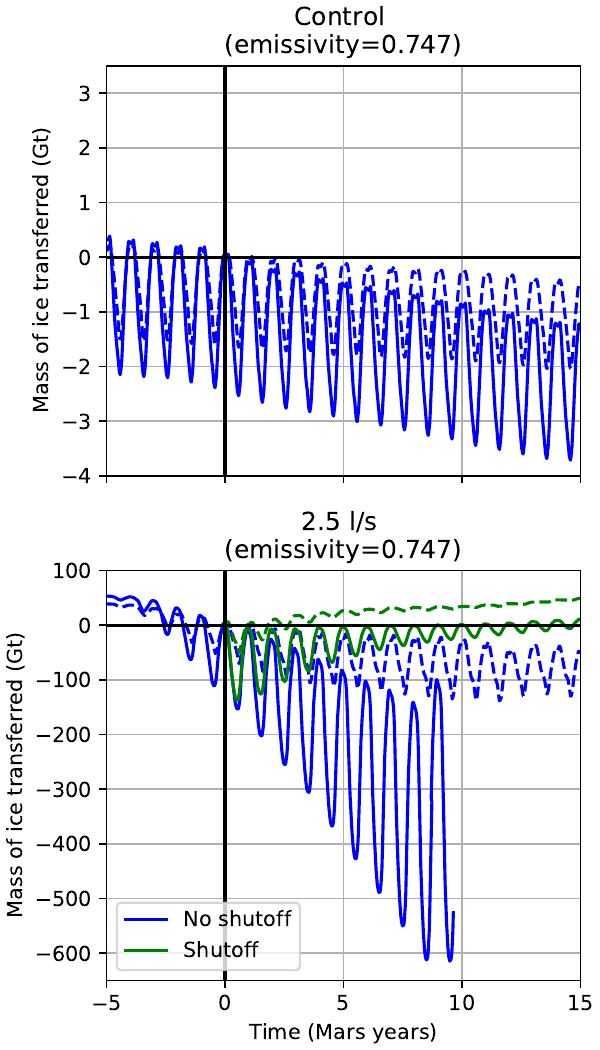}
\includegraphics[width=0.45\textwidth]{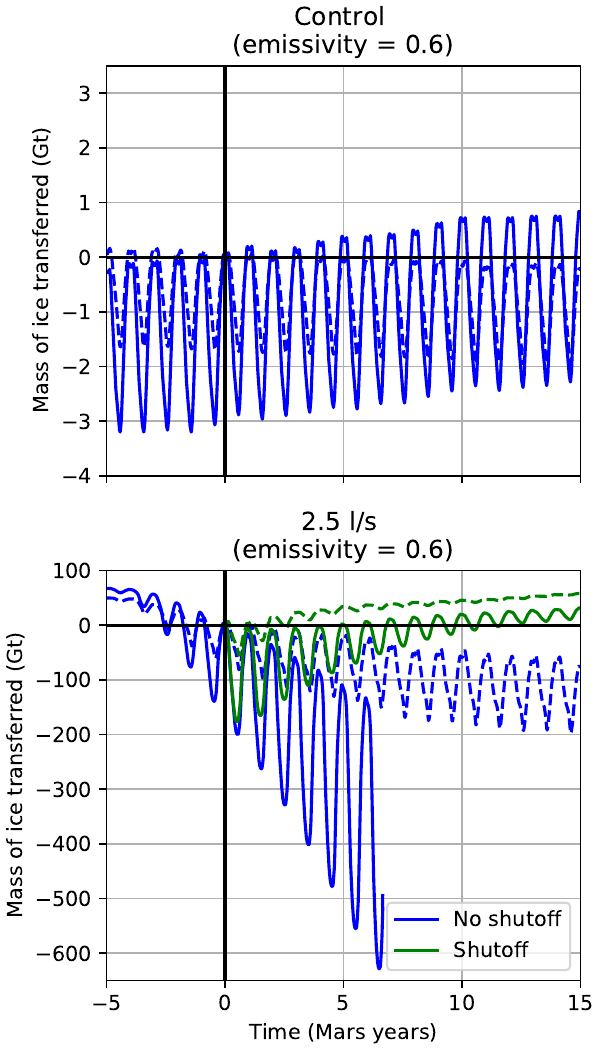}
\caption{Difference between the northern hemisphere and southern hemisphere water inventory in the radiative (solid lines) and non-radiative (dashed lines) cloud feedback scenarios, normalised to the difference when t=0 (5 Mars years after start of aerosol release). Negative values indicate a net transfer of water from the Northern to the Southern hemisphere. Our model shows a small net bias in the control case of <0.1 Gt/year of water transferred from the Northern to the Southern Hemisphere (due to permanent deposition over the South Polar Cap) that can be eliminated or even reversed by reducing the emissivity of South Polar CO\textsubscript{2} ice. However, we model a consistent net transfer of water to the Southern Hemisphere during artificial warming regardless of CO\textsubscript{2} emissivity, that then reverses following shutoff. Our results are therefore robust to these model parameters.}
\label{fig:watertransfer}
\end{figure}

The final major sink of ice is at the South Pole. Approximately 100 Gt of water is seasonally exchanged between the two hemispheres within a single Mars year, with a net transfer of 5-10 Gt of water per Mars year from the Northern to the Southern hemisphere (Fig. \ref{fig:watertransfer}). In the control case, we observe a bias in the hemispheric transfer of water of <0.1 Gt per Mars year, that is dependent on the uncertainty on the emissivity of the South Polar CO\textsubscript{2} cap. This bias disappears during Mars warming, as the annual deposition of water ice at the edge of the residual cap induces an albedo change that causes summer sublimation to be limited by the latent heat of water ice, thereby preventing its complete loss and gradually accumulating over latitudes polewards of 75$^\circ$S to form a perennial ice cap.  Each Southern summer, the northern edge of this new ice cap (75-85$^\circ$S) partially sublimates 10-20 Gt of water vapour into the atmosphere, which contributes to the water cycle as the sublimation of the edge of the North Polar Cap does in Northern Summer.

\begin{figure}[h]
\includegraphics[height=0.38\textheight]{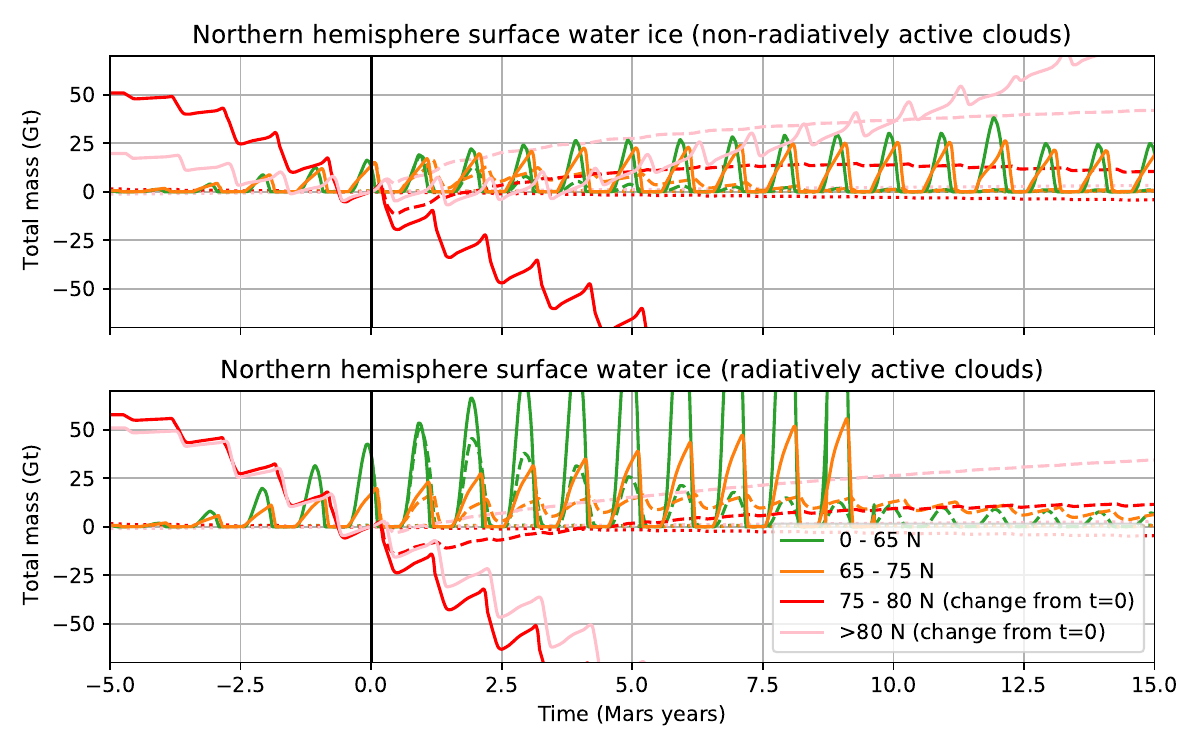}
\includegraphics[height=0.38\textheight]{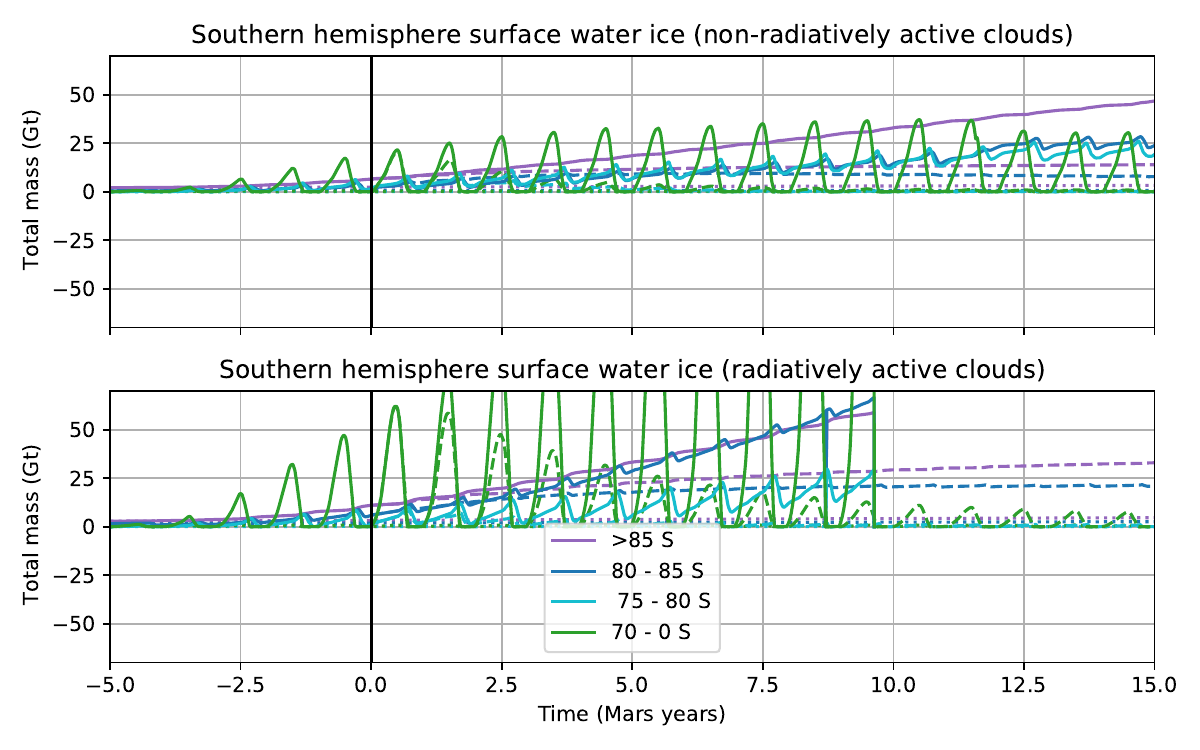}

\caption{\foreignlanguage{american}{During aerosol loading, the South Polar Cap and midlatitude frost increases at
the expense of ice sublimated from the edge of the North Polar Cap, and continues
doing so if shutoff (marked as t=0) does not occur (solid lines). If shutoff occurs
(dashed lines), water sublimates from midlatitude frost and is
redistributed towards both the North Polar Cap and South Polar Cap. When radiative cloud feedbacks are neglected, the thickness of the edge of the South Polar Cap
peaks after 9 Mars years before slowly declining. Radiative cloud feedbacks increase South Polar ice accumulation, with no decline observed within 15 Mars years after shutoff. Control values
are shown in the dotted lines for comparison. Simulations with continuous aerosol release and radiatively active clouds become unstable after 15 Mars years and are therefore not shown after t=9.5 Mars years.}}
\label{reversibilitytimeseries}
\end{figure}

When aerosol release ceases, temperatures drop rapidly and, within a Mars year, the edge of the North Polar Cap (75 - 80$^\circ$ N) is no longer a net source of water vapour in the system. Instead, the edge of the new South Polar Cap (75 - 80$^\circ$ S) gradually sublimates away each summer without being replenished in winter by water sublimating off the North Polar Cap. In addition, the rapid temperature drop results in formerly seasonal frost between 65$^\circ$ and 75$^\circ$ N becoming perennial within the first 5-10 Mars years following aerosol shutoff as summer temperatures are no longer warm enough there to fully sublimate ice that accumulates there in winter. This therefore results in a reversal of hemispheric water vapour transfer, with water sublimating off the edge of the South Polar Cap and from seasonal frost in the mid-latitudes being redeposited back to the North Polar Cap, including in the region between 75 - 80$^\circ$ N where approximately 15-20\% of the initial ice lost is replenished. Although most of this water is permanently redeposited onto the North Polar Cap, the residual South Polar Cap also net accumulates water ice year-on-year (Fig. \ref{reversibilitytimeseries}), with summer sublimation poleward of 85$^\circ$ S being negligible. When radiative cloud feedbacks are not taken into account, we find
that ice between 80$^\circ$ and 85$^\circ$ S peaks in thickness $\sim$6 Mars years after aerosol
shutoff, before the cap slowly declines at a rate modulated by the assumed emissivity of the CO\textsubscript{2} ice deposited there (the lower the emissivity, the more rapid the sublimation). The moister atmosphere induced by radiative cloud feedbacks
accelerates the accumulation of South Pole ice, and even 15 Mars years after aerosol release
has ceased, latitudes poleward of 80$^\circ$ S continues to accumulate ice year-on-year. The water vapour content of the atmosphere
plateaus 10-15 Mars years after shutoff at a new metastable equilibrium
significantly moister than the pre-warming state, since the gradual sublimation of the South Polar Cap between 80 - 85$^\circ$ S and, to a lesser extent, the gradual sublimation of the new perennial ice layer between 65$^\circ$ and 75$^\circ$ N, continues to supply water vapour into the atmosphere decades after aerosol release has ceased. When radiatively active clouds are accounted for, the seasonal pattern of atmospheric water content resembles that of a recent Mars scenario without the residual South Polar CO\textsubscript{2} cap \cite{vos2025} after 15 Mars years, featuring a small spike in global water vapour content during northern summer and a larger spike during the perihelion season corresponding to the seasonal sublimation of northern and southern frost and ice respectively. %, indicating a convergence towards a new metastable climate state in which both polar caps contribute to the Mars water cycle.
Therefore, within our model framework, the change in the main surface ice reservoirs is not
reversible over decadal timescales, and will continue to drive a water
cycle dramatically different from that of present-day Mars for at
least a century. 

\subsection{Effect on the CO\textsubscript{2} cycle}

\begin{figure}[h]
\includegraphics[width=1\textwidth]{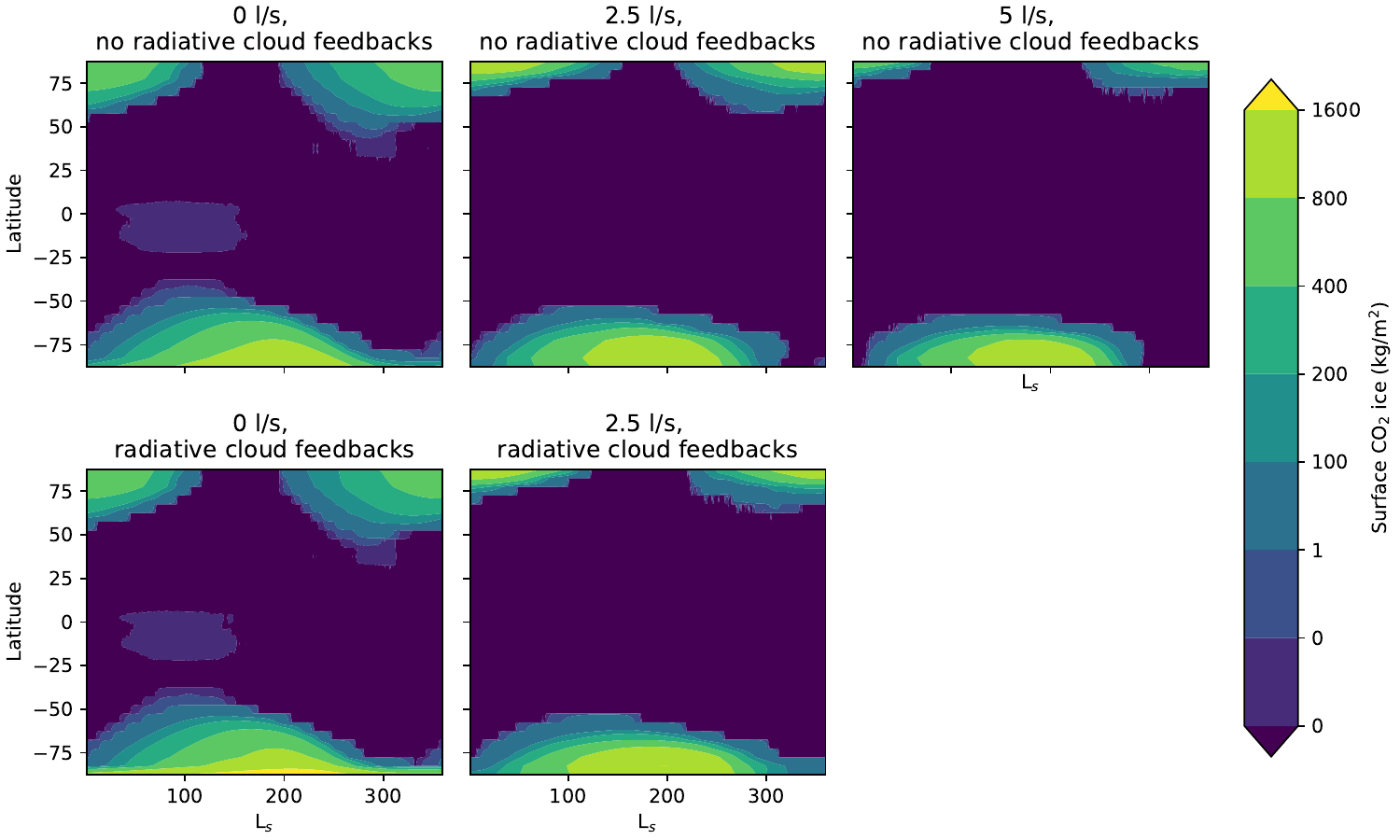}

\caption{\foreignlanguage{american}{Seasonal CO\protect\textsubscript{2} ice condensation continues to
occur over the poles, with artificial warming causing a retreat in
the latitudinal extent of the CO\protect\textsubscript{2} ice cap but an increase in the thickness and seasonal persistence of the CO\protect\textsubscript{2} cap.}}
\label{co2cycle}
\end{figure}

Although the winter latitudinal radius of each cap decreases
by $\sim$10$^\circ$ for an aerosol loading of 2.5 l/s, the
increased atmospheric pressure due to the decreased global CO\textsubscript{2}
ice deposition results in faster deposition of CO\textsubscript{2}
ice over the winter polar cap due to an increase in the frost point of CO\textsubscript{2}
relative to the local surface temperature. The latent heat of this extra CO\textsubscript{2} ice also results in surface temperatures at the poles becoming more resistant to summer warming relative to the control case, especially at the North Pole where surface CO\textsubscript{2} ice is modelled to have a higher albedo than at the South Pole. We observe seasonal CO\textsubscript{2} ice condensation in our model (Fig. \ref{co2cycle}) even in a 5 l/s aerosol loading scenario. 
The Martian CO\textsubscript{2} cycle is therefore highly robust to artificial warming by aerosols, which can create warm regions of Mars while allowing Mars to seasonally condense
out its main atmospheric constituent as before - a feature that makes it unique among the eight IAU-defined planets within the Solar System (setting aside Triton and Pluto; \citeNP{gaoohno2025}).

\begin{comment}
- Hovmoller plot (compare with and without cloud feedback, one during
day and one during night) of surface ice, compare with water vapour
and cloud

- Line plot

- Discussion of icy highlands, map of ice as a function of season?

- Check for snowmelt (where do we get ice/seasonal frost above 273K?
Probably nowhere).

- CO2 cycle

- enforced cold trap at SP.
\end{comment}

\noindent \subsection{Subsurface ice stability}

We will now assess how warming Mars would affect the stability of subsurface ice. Warming subsurface ice could sublimate it, but this could be compensated for by an increased rate of ice formation due to the moister atmosphere. Midlatitude subsurface ice deposits have a distribution that closely corresponds to that expected for equilibrium with the modern climate \cite{mellonjakosky1995}, but are thought to have formed under high obliquities that could take hundreds of thousands of years \cite{laskar2004,smith2020,becerra2021}
to re-occur. To keep midlatitude subsurface ice deposits stable year-on-year, sublimation must be balanced by downward diffusion of water vapour back through the
soil from the atmosphere - in other words, warming of the subsurface must be compensated for by an increase in humidity of the
atmosphere. Moreover, if the midlatitude deposits are to be made habitable, then
some of the more accessible subsurface ice should melt seasonally, and not just sublimate or remain frozen.

Following \citeA{mellon2004} and \citeA{schorghoferaharonson2005}, 
subsurface ice of temperature $T(z)$
beneath a thickness \emph{z} of permeable soil is stable when:

\begin{equation}
\left\langle \frac{min\left(P_{surf}^{H2O},P_{svp}^{H2O}(T_{surf})\right)}{T_{surf}}\right\rangle _{year}-\left\langle \frac{P_{svp}^{H2O}(T(z))}{T(z)}\right\rangle _{year}\geq0
\end{equation}

\noindent where $T_{surf}$ and $P_{surf}^{H2O}$ are, respectively, the temperature
and water vapour partial pressures at the surface, and $P_{svp}^{H2O}$
is the saturation vapour pressure of water at temperature
$T(z)$. If the first term exceeds the second term,
subsurface ice will be deposited at depth $z$ and the ice table will move toward the surface.
In this study, we do not include the effect of pore space
filling of soil with ice, which would increase the thermal inertia 
of the soil \cite{slegler2012,steele2017}. As a result, our model 
understates the stability of ice at depths $<$1 m in the
midlatitudes relative to observed data
\cite{dundas2018,dundas2021,morgan2025}.

We parametrise the stability of subsurface ice for a given soil depth $z$ in both the pre-terraformed control case and the artificially warmed case using a single quantity, $\theta_{z}$:

\begin{equation}
\theta_{z}=atan2\left(s_{warmed},s_{control}\right)
\end{equation}

where:
\begin{equation}
s=\left\langle \frac{min\left(P_{surf}^{h2o},P_{svp}^{h2o}(T_{surf})\right)}{T_{surf}}\right\rangle _{year}-\left\langle \frac{P_{svp}^{h2o}(T(z))}{T(z)}\right\rangle _{year}
\end{equation}

\noindent defines the stability condition at depth $z$ for the two respective scenarios.

The sign of $\sin(\theta_{z})$ gives the net stability
in the warmed case, and the sign of $\cos(\theta_{z})$ provides
the net stability in the control case. Therefore, it is convenient to classify our $\theta_{z}$ results into four separate stability scenarios: a) $0\leq\theta_{z}\leq\frac{\pi}{2}$ (stable ice under both control and warmed conditions), b) $\frac{\pi}{2}\leq\theta_{z}\leq\pi$ (stable ice under warmed conditions, unstable ice under control conditions), c) $-\frac{\pi}{2}\leq\theta_{z}\leq0$ (unstable ice under warmed conditions, stable ice under control conditions) and d) $-\pi\leq\theta_{z}\leq\frac{\pi}{2}$ (unstable ice under both control and warmed conditions). We postprocess MarsWRF output to assess the stability of subsurface ice, if it were present. We do not explicitly model ground ice within MarsWRF.

To calculate how fast heat is conducted through soil using MarsWRF,
we divide the subsurface into 15 homogeneous layers from the surface
to $\sim$15 m depth, with the width of each layer increasing
with depth. The thermal conductivity $k_{reg}$ varies spatially and is derived from Mars Global Surveyor Thermal Emission
Spectrometer (MGS-TES) thermal inertia data \cite{putzigmellon2007} assuming a uniform bulk density
$\rho_{reg}$~=~1500~kg~m\textsuperscript{-3} and specific heat
capacity $c_{p}$~=~837~J~kg\textsuperscript{-1}~K\textsuperscript{-1}.
The highest thermal conductivity values (\textasciitilde 0.5 W m\textsuperscript{-1} K\textsuperscript{-1}
) are found at the poles and the lowest values (\textasciitilde 1 mW
m\textsuperscript{-1} K\textsuperscript{-1}) are in dust
deposits near the equator. We can define a thermal penetration depth $ \approx \sqrt{\frac{k_{reg}}{c_{p}\rho_{reg}}t}$ where $t$ is the duration of surface warming. Ten Mars years of engineered aerosol
release corresponds to the top \textasciitilde70 cm of the surface affected by
artificial warming in low-$k_{reg}$ soils, while in the most conductive regions the warming penetrates down to \textasciitilde
15~m. In most of Mars, however, warming would reach a depth of a few metres
after 10 Mars years. Here we consider only the topmost metre of the soil. This topmost metre is the most accessible
for human use, and can be probed using orbital gamma ray and neutron spectroscopy \cite{morgan2025}.

\begin{figure}[h]
\selectlanguage{american}%
\includegraphics[width=0.55\textwidth]{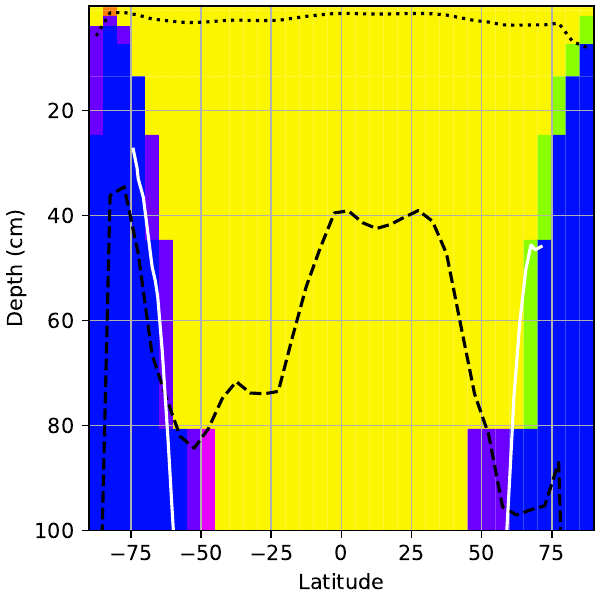}\includegraphics[width=0.45\textwidth]{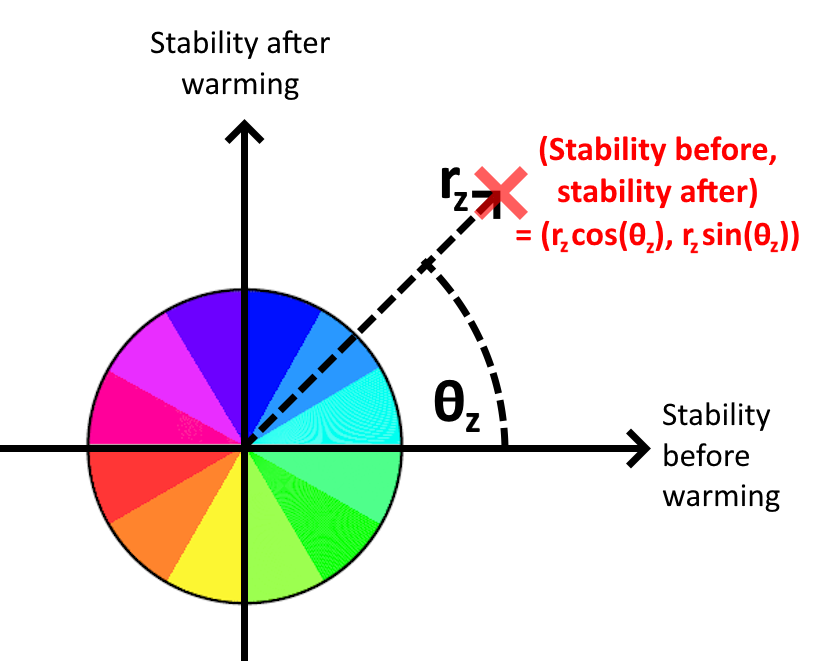}

\caption{Zonally-averaged interannual stability of shallow subsurface water
ice as a function of depth, comparing 10 Mars years of aerosol release with
the control scenario. The dotted and dashed black lines respectively
show the diurnal and seasonal penetration of thermal forcing, while
the white lines show the predicted midlatitude ice deposit extent
assuming no soil pore filling \cite{schorghoferaharonson2005}.
Colours correspond to the value of \textgreek{θ}\protect\textsubscript{z}
(legend on right). Ice is destabilised (green) from the
Northern midlatitude deposit, and ice accumulates (purple and pink)
in the Southern midlatitude deposit and at depths in the Northern
midlatitude deposit less affected by seasonal forcing. Pore ice filling was neglected in this model; in reality the ice table in the mid-latitudes is likely to be shallower than our model predicts.}
\label{fig:shallowice}\selectlanguage{english}%
\end{figure}

Fig. \ref{fig:shallowice} shows how subsurface ice stability shifts after 10 Mars years of artificial warming
(2.5 l/s scenario, radiatively active clouds, neglecting pore closure by ice formation). Our
control (0 l/s) scenario results are in line with those of \citeA{schorghoferaharonson2005}
for zero pore space filling, with stable subsurface ice in the topmost metre of regolith mostly confined to latitudes poleward of 60$^\circ$ north and south. At the poles, artificial warming acts to increase the stability of subsurface ice since the order-of-magnitude increase in surface water vapour concentration due to increased summer sublimation of water ice dominates over the increase in temperature in the shallow subsurface, which is buffered by the latent heat of the overlying ice cap. At lower latitudes, on the other hand, the increase in shallow subsurface temperatures dominates over the increase in surface humidity due to artificial warming, and so any shallow subsurface ice that exists there is rendered even less interannually stable. We also observe a hemispheric asymmetry where colder winters and more persistent surface CO\textsubscript{2} ice at southern high latitudes result in subsurface ice increasing in stability at the edge of the South polar ice cap following artificial warming, but decreasing at the edge of the North polar ice cap down to a depth of \textasciitilde80 cm, although we would likely observe stable ice at shallower depths in both hemispheres if subsurface ice pore occlusion was taken into account in our model. Eventually, the water cycle will reach
equilibrium, with the upper parts of the North Polar Layered Deposit and the edges of the Northern
midlatitude deposits permanently destabilised.

As warm-season average temperatures remain below freezing at latitudes with stable subsurface ice in our model, MarsWRF does not output temperatures
above freezing at locations where ground ice is predicted to be stable, even in the highest aerosol loadings
studied here. In reality, however, subsurface ice is observed on Mars at $<1$~m depth poleward of 50-55$^\circ$ latitude, including in the Southern Hellas
basin \cite{dundas2021} where we model warm-season average temperatures of $>$273 K with daytime peaks $>$320 K. This daytime warming would sublimate any
water ice within the top $\sim$10 cm of soil, while deeper
ice would be less sensitive to diurnal temperature forcing.
The cooling effect of latent fluxes from the
sublimation of water ice \cite{langeforget2025} is muted
for subsurface ice deeper than $\sim$50 cm (the level at which
we estimate the evaporitic cooling rate to be equal to 10 W m\textsuperscript{\selectlanguage{american}%
-2} for ice sublimating at 273 K).

\noindent \section{Model limitations}

In this paper, we present the results of the first simulation of water cycle feedbacks on an artificially warmed Mars. However, our model is limited by a number of variables that require the following future measurements to constrain:

\begin{itemize}
\item Warming is sensitive to dry deposition rates for sub-$\mu$m particles, for which empirical data in Mars-like conditions is lacking.
\item Maps of water ice and subsurface CO\textsubscript{2} remain incomplete for present-day Mars \cite<e.g.>{jakoskyedwards2018,Broquet2020,buhlerpiqueux2021,jakoskyhallis2024}. %, as does the role of the soil and near-subsurface in the water cycle.
\item We cannot predict the present-day Martian dust cycle, including large storms \cite{kahre2017}. To do this requires more data on dust particles, dust lofting, and dust sources and sinks. These unknowns make it hard to know how the dust cycle would change if Mars was artificially warmed.
\end{itemize}

Ice sampling for isotopic data would be needed to quantify the water cycle feedback on temperature on Mars during historical periods of high obliquity, which still remains unknown \cite<e.g.>{Mischnarichardson2005} as they are highly dependent on model parametrisations of the water cycle. High-obliquity climate states of recent Mars may be the best (though imperfect) analogue to the early stages of artificial Mars warming due to the increased sublimation of water vapour induced by direct solar irradiation of the polar caps, and would therefore allow for further constraints on how the Martian climate system behaves under high atmospheric loadings of water.

Our results would also depend on a number of properties of the engineered aerosol itself, which would have to be measured experimentally. One is the ability of engineered aerosol particles to either self-agglomerate or to stick to natural dust. Another is their potential to nucleate water ice cloud particles. This could dramatically change Mars cloud properties. Because these measurements have not been conducted, we ignored them in this study, but this could lead to incorrect predictions of the effect of radiative cloud feedbacks on a warmed Mars \cite{madeleine2012,navarro2014,kahre2017mamo}. Aerosols settling on the surface could also change the surface albedo, and thereby alter surface temperature and/or water sublimation rate. By assuming that particles are irreversibly removed by dry deposition, we neglected other factors that might alter particle lifetime, such as (a) degradation of particles, (b) re-lofting of deposited particles, which could lead to a longer effective aerosol lifetime (analogous to that of the present-day dust cycle), and (c) agglomeration of particles causing faster sedimentation. If particles neither degrade nor agglomerate, and are easily re-lofted, then Mars would stay warm indefinitely even if particle release is cut off. 

Finally, we neglected the self-stabilising effects on subsurface ice of both soil pore filling (which raises thermal inertia) and evaporitic cooling. Both effects lower peak ice temperatures and so reduce sublimation.

Even with better measurements of Mars and better lab data, there are basic limits to the predictive power of models of planetary climate change, as there are with even our best models of Earth's climate response to ongoing human-induced warming \cite{Shaw2025}. As warming Mars by 30~K would be a big change to Mars' energy balance, assumptions that only apply to the modern climate might lead to incorrect extrapolations in predicting the warmed climate. Thus, any intentional alteration of Mars' surface temperature should ideally start small, start local, and be equipped with the ability to course-correct \cite{macmartinkravitz2019engineering,kitewordsworth2025}. Future tests could assess if aerosol warming can be confined to a small topographic basin, for instance, by releasing aerosol particles from the centre of the Hellas Basin that are engineered to have a short lifetime in the atmosphere that does not allow global dispersal.
 
%These uncertainties highlight that there is much that
%is still to be learnt about the modern-day Martian environment before
%we can even be close to determining the long-term effects of artificial
%warming on the Martian climate, what elements of the Martian environment
%will be irreversibly altered that make it unique within the Solar
%System and would prove of major scientific value, 

%Several modelling approaches to warm Mars in a controlled manner without the global redistribution of ice could be investigated. 

In summary, processes left unaccounted for in existing models of a warmed Mars could lead to undesirable outcomes. We do not yet know whether warming Mars would permit a new biosphere or aid human exploration. % The largest second-biosphere experiment to date, Biosphere 2, did not demonstrate a self-sustaining ecosystem. 
Thus, more data (including, but not limited to, Mars sample return) is needed about Mars to enable informed decisions about the role of humans in Mars' future and the possibility of permanent inhabitation \cite{kitewordsworth2025,NASEM2025}.

\noindent \subsection{Conclusions}

\begin{sidewaysfigure}
\selectlanguage{american}%
\includegraphics[width=1\textwidth]{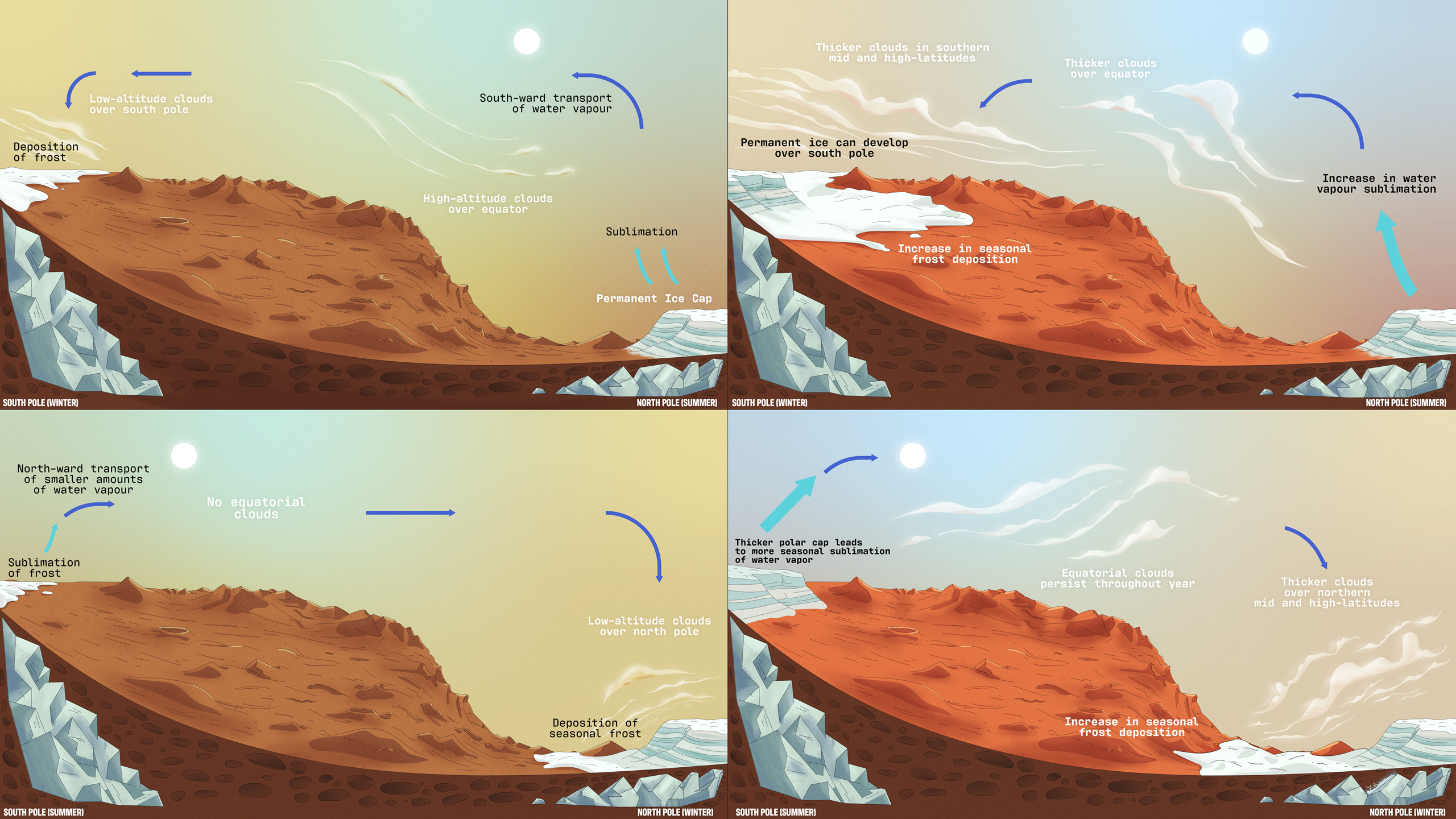}

\caption{We summarise differences in the Martian water cycle between the existing modern Mars case (left) and the artificially warmed case (right), notably a) the increase in cloud cover and b) the increased participation of the South Pole in the water cycle. \textit{Figure by Vlad Socianu}}
\label{fig:vladwatercycle}\selectlanguage{english}%
\end{sidewaysfigure}

In principle, Mars might be warmed using aerosols \cite{ansari2024}, gases \cite{marinova2005}, or solar reflectors \cite{Handmer2024}. However, the costs, benefits, and possible risks of warming Mars remain unclear.
%Recent studies by \citeA{ansari2024} and \citeA{richardson2025}
%suggested that the release of artificial `engineered' aerosols into
%the Martian atmosphere may allow Mars to be warmed sufficiently to allow for regions of seasonally
%stable liquid water, consistent with earlier studies of warming using super-greenhouse gases %\cite{marinova2005}. 
Previous work neglected the effects of warming on the Mars water cycle, whose radiative impacts could accelerate or delay warming, and move ice to high-altitude zones that are unfavorable for melting.
%changes
%to the Martian water cycle induced by engineered aerosol release on
%the Martian climate system and the Martian environment as a whole.
We modelled the effect on the water cycle of the release of
an engineered aerosol with idealised Mars-warming radiative properties into Mars' atmosphere, neglecting microphysical interactions between aerosols. 
%studying the surface temperature response
%induced both directly by the aerosol release, and indirectly through
%the water vapour radiative feedback and cloud radiative feedback
%(neglecting the effects of supersaturation). 
%In particular, we analysed the sensitivity of our results
%to a) aerosol loading, b) the susceptibility of engineered aerosols
%to being removed from the atmosphere through dry deposition, and c)
%the albedo of the North Polar water ice cap. 
%We also studied the long-term
%impact and reversibility of the release of engineered aerosols on
%the Martian climate system and the distribution of water ice on Mars. 
Within our model framework, we found that a warmer Mars results in a moister atmosphere, with strong radiative cloud feedbacks: nighttime warming over most of the planet, and strong daytime cooling in the winter midlatitudes. Water ice is slowly lost from surface and subsurface reservoirs near the North Pole, and accumulates over the South
Pole. This causes a positive feedback where seasonal sublimation from the South Polar Cap contributes more to the water cycle, boosting the amount of atmospheric water vapour and clouds (see Fig. \ref{fig:vladwatercycle} for a graphical summary). 
In turn, this boosts radiative cloud feedbacks, warming most of Mars but with strong winter cooling
over the midlatitudes. The effects of this change in the water cycle
persist for at least decades even if aerosol release is ended, as the long-term redistribution of ice towards the South Pole 
causes winter radiative cloud feedbacks to cause strong cooling in both the Northern and Southern mid-latitudes, increasing the formation of winter frost that acts as an additional major reservoir of ice in the Martian climate system.
%The longer the period of engineered aerosol release
%into the atmosphere, the greater and more long-lasting the impact on
%the Martian climate system. 

In closing, we emphasise that, even if Mars' surface can be warmed, the planet would still remain a much more hostile environment than the Earth for even the hardiest life forms, let alone humans, and no known extremophile can tolerate all of the stresses of the Martian environment \cite{Coleine2025,debenedictis2025}. Mars lacks a breathable atmosphere and an effective shield against harmful UV radiation, while its soil is salty and contains perchlorates that can inhibit plant growth. 
Warming Mars will therefore not be a substitute for protecting the environment of the only habitable planet we will have for the foreseeable future (the Earth), while changing the Martian environment in ways that could take at least decades to reverse.

\begin{comment}
Better microphysics, inclusion of dynamic dust, re-lofting of nanoparticles,
dry deposition, effect of nanoparticles on albedo of water ice, confineability
with larger particles, clumping, lag deposition over the North Polar Cap, sublimation
of CO2 deposits in South Pole.
\end{comment}

\begin{acknowledgments}
This work was partly funded by a Residency  for E.S.K at Astera Institute. This work used resources from the University of Chicago's Research Computing Center and the NASA High-End Computing (HEC) Program
through the NASA Advanced Supercomputing (NAS) Division at Ames Research
Center. A portion of this work was conducted at the Jet Propulsion Laboratory, California Institute of Technology, under contract with NASA. We thank the PlanetWRF development team for their support.
\end{acknowledgments}

\noindent \subsection*{Open access statement:}\selectlanguage{american}%

All model output is available on Zenodo (doi:10.5281/zenodo.18717694), including input files that allow for reproducibility of all of the figures presented here. Requests for the MarsWRF source code should be submitted to mir@aeolisresearch.com.

\selectlanguage{american}%
\bibliography{mars}

\selectlanguage{english}%

\end{document}